\documentclass[a4paper,fleqn]{cas-dc}

\usepackage[authoryear]{natbib}
\usepackage{booktabs}
\usepackage{multirow}
\usepackage{xspace}
\usepackage[most]{tcolorbox}
\usepackage{enumitem}
\usepackage{tikz}
\usetikzlibrary{arrows.meta,positioning,fit,backgrounds}

\begin{document}
\let\WriteBookmarks\relax
\def\floatpagepagefraction{1}
\def\textpagefraction{.001}

\shorttitle{Cross-Project Bug Localization: What Works, What Doesn't, and Why}
\shortauthors{Rao and Chimalakonda}

\title[]{What Works, What Doesn't, and Why: An Empirical Investigation of Cross-Project Bug Localization}

\author[1]{A Eashaan Rao}[orcid=https://orcid.org/0000-0001-8499-7538]
\cormark[1]
\ead{cs21d002@iittp.ac.in}

\author[1]{Sridhar Chimalakonda}[orcid=https://orcid.org/0000-0003-0818-8178]
\ead{ch@iittp.ac.in}

\affiliation[1]{organization={Indian Institute of Technology Tirupati},
            city={Tirupati},
            postcode={517619},
            state={Andhra Pradesh},
            country={India}}

\cortext[1]{Corresponding author}

\begin{abstract}
Bug localization models require substantial within-project training data that newly
active or small open-source projects rarely have. Cross-project learning (CPL) promises
to close this gap by reusing labeled bugs from mature projects, but prior CPL studies
evaluate a single model against ample-data within-project baselines that data-scarce
projects cannot reach. We ask whether CPL's effectiveness is a property of the transfer
setting or depends on model architecture. We evaluate three architecturally distinct
models, embedding-based BLAZE, convolutional TRANP-CNN, and GNN-based adversarial COOBA,
on 63 source$\to$target pairs from 13 open-source Python projects under identical bug
corpora, data splits, retrieval candidates, and baselines. The primary baseline is
within-project limited training (WP-small, $\sim$20\% of target labels, the same
target-side share as CP-transfer). CPL outperforms WP-small most clearly for TRANP-CNN
(Cohen's $d = 0.95$, win rate 81.0\%, $p < 0.001$, robust to clustering pairs by target
project), and less so for BLAZE ($d = 0.49$, win rate 60.3\%, $p = 0.002$; not
significant once pairs are clustered by target). COOBA, under our re-implementation,
shows a small gain that does not survive multiple-comparison correction ($d = 0.27$,
$p = 0.027$). Negative transfer affects 9.5\% of TRANP-CNN pairs and 30.2\% of BLAZE's
and COOBA's. Cross-model direction agreement is only 44--54\%: the same pair can help
one architecture and harm another, although the models' gains are positively correlated
($\rho = 0.24$--$0.42$). CPL outcomes are therefore architecture-dependent:
practitioners should validate CPL for their chosen model, and CPL studies should report
results per architecture.
\end{abstract}


\maketitle

\section{Introduction} \label{sec:intro}

Bug localization, automatically identifying the source location(s) responsible for a reported defect, is a labor-intensive activity in software maintenance. Deep learning-based models have demonstrated strong performance on established benchmarks~\citep{lam2017bug,liang2022modeling,ciborowska2022fast,hossain2024deepdive,zhu2025funloc}, but their effectiveness depends on a readily-overlooked prerequisite: substantial labeled training data (paired bug reports and ground-truth source files) that a project accumulates only through years of sustained, tracked issue resolution.

For many projects this accumulation never reaches sufficiency or starts from zero. Early-stage or infrequently-reported projects may have only dozens to a few hundred verified mappings exactly when a localization tool is most needed. The difficulty is not merely data volume: developer familiarity with a codebase materially affects debugging efficiency~\citep{wang2020familiarity}, so a transferred model must compensate for exactly the contextual knowledge that familiarity provides. The scale is illustrated by the project with our smallest bug corpus, \texttt{pydata/xarray}, which has only 122 verified mappings. A standard limited-data split (20\% of the corpus) yields just 24 labeled training examples, too few for reliable generalization and representative of any project in its first years of bug tracking, precisely where effective localization would be most valuable.

Cross-project learning (CPL) is a natural response: train on a data-rich source project, then apply or fine-tune on the data-scarce target, on the intuition that the bug-report-to-code relationship may transfer across projects sharing a language or coding style. Several models demonstrate this, including TRANP-CNN~\citep{huo2019deep}, COOBA~\citep{zhu2020cooba}, BL-GAN~\citep{zhu2022bl}, and BLAZE~\citep{chakraborty2025blaze}. Yet a fundamental question remains unanswered: \textit{does CPL effectively help the practitioners it is designed to serve?} Each of these studies shows its model outperforming an ablated baseline or earlier CPL method, but their within-project comparators use the \emph{majority} of the target's labeled history; data-scarce practitioners cannot match this volume, so these experiments show that CPL surpasses a standard they cannot reach, not whether it beats the limited training they \emph{can} perform. In the analogous task of cross-project defect prediction, a 90-study survey found transfer context-dependent, helping in some conditions and hurting in others with no single reliable predictor~\citep{hosseini2017systematic,zhang2022negative}. Whether CPL helps teams with limited within-project data is, to our knowledge, unanswered in the bug localization literature.

A second gap compounds this: to our knowledge, every published CPL paper evaluates a single model of its own design, so it is unknown whether demonstrated benefit is a property of the cross-project paradigm or an artifact of one model's inductive biases. A practitioner reading one CPL paper and choosing a different tool cannot know whether its findings apply. No existing study uses multiple CPL models as independent instruments across the same project pairs.

We fill both gaps with a controlled empirical study across three architecturally distinct, well-established CPL models spanning the dominant paradigms (embedding-based BLAZE, GNN-based COOBA, CNN-based TRANP-CNN) on 63 directional source$\to$target pairs from 13 open-source Python projects in four domains (Data Science/ML, Developer Tools, Systems/Cloud, Applications), under four scenarios isolating the contribution of cross-project knowledge. The primary baseline throughout is \emph{within-project limited training}, the data a newly active project actually has, not prior work's ample-data ceiling. A core methodological contribution, independent of empirical findings, is the evaluation protocol itself: since each prior paper proposes its own architecture, corpus, and comparator, cross-paper comparisons conflate architectural differences with varying conditions, making it difficult to attribute outcomes to any specific cause. Here, all three models share the same 63 pairs, bug reports, train/test splits, FAISS indexes, and primary baseline, treated as \textbf{independent scientific instruments under an identical experimental protocol}, supporting attribution of outcome differences to architecture rather than incidental conditions, though not by itself establishing the causal mechanism behind any given outcome.

\begin{tcolorbox}[colback=white, colframe=black, float=b, title=\textbf{Contributions and Summary of Findings}]
\textbf{What we contribute.}
\textbf{(C1)} A controlled multi-instrument evaluation framework: three architecturally
distinct CPL models evaluated under an identical protocol; same 63 source$\to$target
pairs, bug reports, data splits, FAISS indexes, and primary baseline. To our knowledge,
no prior CPL study provides this.
\textbf{(C2)} To our knowledge, the first systematic comparison of CPL against within-project
\emph{limited} training: the data a newly active project actually has, not the
ample-data ceiling used in prior work.
\textbf{(C3)} Negative-transfer rates for all three architectures under
identical conditions: 9.5\%/30.2\%/30.2\% of pairs for
TRANP-CNN/BLAZE/COOBA against a budget-matched WP-small baseline
(Section~\ref{sec:results}), the first such rates reported for CPL bug
localization.

\textbf{Summary of findings.} CPL benefit varies across the three
models tested on identical data: a large effect for TRANP-CNN, a smaller one for
BLAZE, and a small effect for COOBA that does not survive multiple-comparison
correction; Section~\ref{sec:results} reports the full per-research-question
breakdown (win rates, effect sizes, negative-transfer risk, cross-model
agreement). The divergence in zero-label (cold-start) behaviour is consistent
with whether a model's ranking mechanism requires target-side calibration at
inference time, though BLAZE also reranks with the same embedding signal that
built the candidate pool, a non-exclusive alternative this study cannot
separate; Section~\ref{sec:architecture} develops both as interpretations
rather than isolated causes.
\end{tcolorbox}

Answering this rigorously presents three challenges. \textbf{(Ch1)} Attributing outcomes to architecture requires eliminating heterogeneity in training data, evaluation corpora, and baselines; to our knowledge, no prior CPL study has enforced this. \textbf{(Ch2)} Choosing the baseline is conceptually subtle: data-scarce practitioners cannot reach prior work's ample-data ceiling, so any comparison against it answers a different question from the one practitioners face. \textbf{(Ch3)} Assembling a corpus large enough for statistical power across diverse pairs requires a principled pair-selection strategy, since a na\"{i}ve all-pairs design would exceed any realistic compute budget.

The rest of the paper is organized as follows. Section~\ref{sec:related} reviews related work on bug localization and CPL. Section~\ref{sec:design} describes the study design. Section~\ref{sec:results} presents results for each research question. Section~\ref{sec:discussion} discusses practical implications and threats to validity. Section~\ref{sec:conclusion} concludes.

\section{Background and Related Work}
\label{sec:related}

\subsection{Bug Localization}

Bug localization ranks source files (or functions) by their likelihood of
containing the defect described in a natural-language bug report. Early
approaches relied on information-retrieval similarity between bug report
text and source file content~\citep{zhou2012should,saha2013improving,ye2014learning},
using measures such as BM25~\citep{robertson2009probabilistic} or VSM over
code tokens: simple and training-free, but limited by vocabulary
mismatch between bug report prose and code identifiers.

Deep learning improved on IR baselines by learning joint representations
of bug reports and code: \citet{lam2017bug} (DNNLOC) combined textual
similarity, file change history, and structural features in a multi-layer
network, improving over IR on Java benchmarks; \citet{liang2022modeling}
extended this to function-level localization via a call-dependency graph
neural network; \citet{ciborowska2022fast} replaced conventional encoders
with pre-trained BERT~\citep{devlin2019bert} representations over code
changesets, showing that transfer from large pre-trained models reduces
the labeled-data requirement. A recent survey~\citep{niu2025deep} confirms
labeled within-project data remains the standard assumption across this
literature. The \citet{lee2018bench4bl} benchmark (51 Java projects) is a
reference point for comparing methods~\citep{bogomolov2024long}, though a
systematic review of debugging benchmarks cautions that benchmark choice
materially affects reported performance and cross-study
comparability~\citep{hirsch2022benchmarks}. Careful pre-/post-processing
(stack-trace filtering, stop-word removal, code-structure weighting) can
itself yield substantial gains independent of the underlying ranking
model~\citep{kim2022digbug}.

A consistent limitation of all within-project approaches is assuming the
target has accumulated sufficient labeled bug reports for training, an
assumption that fails for new or data-scarce projects, the exact regime
motivating this study.

\subsection{Cross-Project Bug Localization}

Cross-project bug localization (CPL) addresses data scarcity by
transferring knowledge from a source project to a target. \citet{huo2019deep}
proposed TRANP-CNN, an early deep learning CPL model, a convolutional
network learning shared bug-report--code representations from a source
project and applying them to a target, but evaluated only against an
ample-data within-project baseline, leaving the practically relevant
data-scarce comparison unanswered~\citep{zhu2021trobo}. \citet{zhu2020cooba}
introduced COOBA, using adversarial domain adaptation~\citep{ganin2016domain}
via a gradient reversal layer to learn domain-invariant AST-structured code
representations. \citet{zhu2022bl} proposed BL-GAN, a
semi-supervised approach augmenting scarce target labels with generative
adversarial networks.
\citet{chakraborty2025blaze} proposed BLAZE, a dual-encoder contrastive
model using dynamic chunking and hard-negative mining, evaluating
cross-project and cross-language transfer but testing only a single
architecture, leaving cross-model CPL consistency
unknown~\citep{chandramohan2024supporting,li2025knowledge,reddy2025swerank}.

\looseness=-1 Across these four papers, the gap identified in
Section~\ref{sec:intro} holds without exception: none uses multiple CPL
models as independent instruments, evaluates against the
practitioner-relevant limited baseline, or tests cross-model consistency of
findings.

A distinct emerging paradigm employs large language models in agentic
workflows, such as SWE-agent~\citep{yang2024sweagent} and
Agentless~\citep{xia2024agentless,zhang2024autocoderover,hossain2024deepdive},
bypassing labeled-data requirements by leveraging pre-training directly for
fault localization and patch generation. The relationship to the CPL setting studied here is a task difference, not
a deployment-scenario one: agents select and patch a small shortlist of
files for a specific issue, whereas this study measures fine-grained
ranking across an entire codebase (Top-1, Top-5, Top-10, MRR, MAP), and
agent evaluations do not report file-localization accuracy at these
metrics, so no direct comparison is available or attempted here. Our
corpus also overlaps substantially with agent benchmarks (11 of 13
projects drawn from SWE-bench, Section~\ref{sec:dataset}), so this study
does not itself demonstrate a setting complementary to agents;
privacy-constrained deployment, offline inference, and cost-sensitivity
motivate supervised CPL as a separate line of work. The conditions
associated with CPL success here, chiefly target label volume and
inference-time adaptation mechanism, may nonetheless inform how agents
prioritize code during fault localization.

These claims are made relative to a targeted search rather than a systematic
review. We seeded from four supervised deep-learning CPL bug-localization
papers (TRANP-CNN, COOBA, BL-GAN, BLAZE), selected as the supervised
deep-learning models in this line of work that evaluate cross-project
transfer. We performed backward and
forward snowballing on each seed paper, ran keyword queries combining
cross-project and cross-language with bug localization and fault localization
on the ACM Digital Library and IEEE Xplore, and cross-checked coverage against
a recent survey of deep-learning IR-based bug localization
\citep{niu2025deep}. We screened for work that trains a localization model on
one project's labeled bugs and evaluates it on another. Within that set, none
of the four properties above is satisfied by a prior study. We may have
missed work outside these venues or using different terminology.

\subsection{Transfer Learning in Software Engineering}

The broader literature on cross-project transfer in software engineering
provides useful context~\citep{pan2010survey}. Cross-project defect
prediction (predicting module defectiveness using another project's
training data) has been studied systematically for years:
\citet{hosseini2017systematic} survey 90 such studies and find transfer
context-dependent, helping in some conditions and hurting in others with no
single reliable predictor~\citep{sottomayor2024survey}. Proposed remedies
span feature-level transfer, e.g.\ matching source/target feature
distributions before training a cross-company classifier~\citep{yu2017feature},
and adaptive instance weighting accounting for feature and class
distribution shift jointly~\citep{tong2023array}, though a systematic
review finds model architecture interacts with these strategies in ways
not yet well characterized~\citep{giray2023deep}, a gap this study
addresses directly for bug localization, motivating our characterization
of CPL success conditions and focus on negative transfer as a deployment
risk. The parallel is informative: like defect prediction, bug
localization is a within-project learning task where CPL is possible but
not guaranteed beneficial.

\looseness=-1 The theoretical basis for COOBA's adversarial domain
adaptation is domain-adversarial neural networks~\citep{ganin2016domain},
which minimize feature-level source-target divergence by training a domain
discriminator with reversed gradients during feature extraction, widely
applied in software engineering transfer settings, though its
effectiveness for cross-project code understanding depends on whether the
feature extractor captures genuinely domain-invariant representations, an
assumption we test empirically.

\subsection{Pre-trained Code Embeddings: Design Motivation}

Recent advances in pre-trained code embeddings have raised representation
quality for both within- and cross-project
models~\citep{feng2020codebert,guo2021graphcodebert,wang2021codet5,guo2022unixcoder}.
This study uses BAAI/bge-code-v1~\citep{bge_code}
(Section~\ref{sec:preprocessing}) as a shared first-stage retrieval
backbone across all three models, so the reranked candidate pool is
identical and reranking differences are attributable to the models
themselves. BLAZE additionally uses BGE as its reranking backbone, while
TRANP-CNN and COOBA rerank with their original designs' input
representations (learned token embeddings and GloVe-300 with AST graphs,
respectively; Section~\ref{sec:models}).

\looseness=-1 Together, these four subsections frame the design choices in
Section~\ref{sec:design}: the shared BGE retrieval stage controls the
candidate pool, the three architecturally distinct models serve as
independent instruments for testing CPL consistency, and the limited
within-project baseline addresses the evaluation gap identified above.

\begin{figure*}
    \centering
    \includegraphics[width=0.75\linewidth]{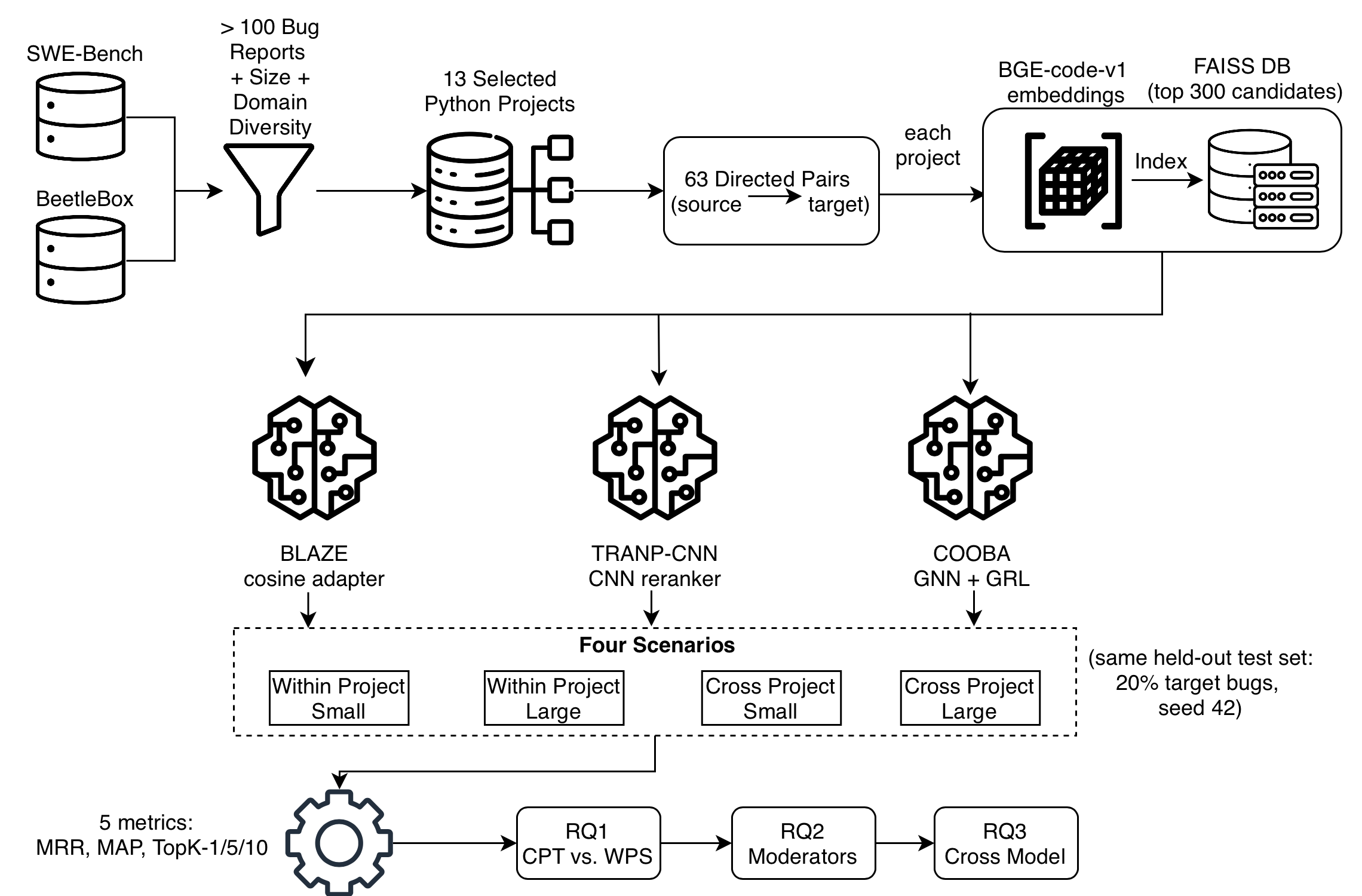}
    \caption{End-to-end study design. Python projects from SWE-bench and BeetleBox are filtered by bug-report volume and diversity criteria to yield 13 projects, which are paired into 63 directed pairs via structured selection. All target projects share one BGE-code-v1 embedding and one FAISS top-300 retrieval stage (the controlled backbone). Three architecturally distinct models (BLAZE, TRANP-CNN, COOBA) rerank the same candidate pool in four training scenarios, with the held-out test set held constant throughout. The shared candidate pool and held-out test split support comparison across the three implementations; the training scenarios differ in source data and target-label budget.}
    \label{fig:study_design}
\end{figure*}

\section{Study Design}
\label{sec:design}

Figure~\ref{fig:study_design} summarizes the end-to-end design. The study is
organized around a single methodological principle: making CPL benefit
attributable to the cross-project paradigm and to model architecture, rather
than to incidental differences in training data, candidate pools, or
evaluation conditions. Every design choice below, the shared retrieval
backbone, identical datasets and held-out splits, the four-scenario
training framework, the primary baseline, and the structured pair
selection, follows from and is justified by this principle.

A \textbf{FAISS IndexFlatIP}~\citep{johnson2019billion} retrieval stage,
built from identical BGE embeddings across all target projects, serves as
the first-stage candidate pool shared across all three models, treated as
\textbf{independent scientific instruments} under an identical protocol: the
same projects, bug reports, train/test splits, FAISS retrieval indexes,
evaluation metrics, and primary baseline. As argued in
Section~\ref{sec:intro}, this supports attributing outcome differences to
architecture rather than experimental conditions, without by itself
establishing mechanism.

\paragraph{Reporting standards.} This study follows the ACM SIGSOFT Empirical
Standards~\citep{ralph2021empirical}, specifically the General Standard and the
Data Science Standard. A completed checklist against both standards is included
in the replication package (see Data Availability).

\subsection{Research Questions}

This study is organized around three research questions, each targeting a distinct
aspect of CPL viability.

\begin{description}
  \item[\textbf{RQ1}] \emph{Does cross-project learning outperform within-project
    training under label scarcity?} Tests the core CPL premise against the
    practitioner-relevant baseline (Section~\ref{sec:intro}): the limited
    within-project data a project actually has, not prior work's ample-data
    ceiling.

  \item[\textbf{RQ2}] \emph{What project-level characteristics determine whether CPL
    succeeds or fails?} Even if beneficial on average, practitioners need to
    know \emph{when} CPL is and is not appropriate: moderating conditions
    (target size, domain alignment) and the risk of negative transfer.

  \item[\textbf{RQ3}] \emph{How consistent are CPL outcomes across different model
    implementations?} A paradigm-level benefit should be consistent across
    architectures; a model-specific one would not generalize. Also asks
    whether project metadata reliably guides source selection.
\end{description}

\paragraph{Hypotheses and motivated expectations.}
RQ1 is confirmatory, a directional claim against a quantifiable baseline
under a controlled protocol, so we state formal hypotheses. They compare
CP-transfer (CPT), trained on a source project's bugs plus $\sim$20\% of the
target's, with WP-small (WPS), trained on the same $\sim$20\% of target bugs
alone (both defined in Section~\ref{sec:scenarios}). For each evaluation
metric $M \in \{\mathrm{MRR}, \mathrm{MAP}, \text{Top-1}, \text{Top-5},
\text{Top-10}\}$ (Section~\ref{sec:metrics}):
\begin{description}
  \item[$H_0$:] CP-transfer does not outperform within-project limited training:
        $M_{\mathrm{CPT}} \leq M_{\mathrm{WPS}}$.
  \item[$H_1$:] CP-transfer outperforms within-project limited training:
        $M_{\mathrm{CPT}} > M_{\mathrm{WPS}}$ (one-sided).
\end{description}
The five tests are corrected together with a Bonferroni adjustment
(Section~\ref{sec:statmethods}). Where a single summary figure is needed
(effect sizes quoted in the text, the negative-transfer definition, and the
RQ2 and RQ3 analyses), we report MRR, the summary metric most widely used in
bug localization; Table~\ref{tab:wilcoxon} gives all five.
This direction is motivated by CPL literature demonstrating transfer
feasibility~\citep{huo2019deep,chakraborty2025blaze,zhu2020cooba}, though all
prior comparisons use ample-data baselines; whether $H_1$ holds against the
baseline data-scarce practitioners can actually reach is precisely what has
not been tested.

RQ2 and RQ3 are exploratory, so we state literature-grounded expectations
rather than formal hypotheses. For \textbf{RQ2}, cross-project defect
prediction research finds transfer success context-dependent with no single
reliable predictor~\citep{hosseini2017systematic}; we expect analogous
conditions for CPL but make no a priori prediction about which moderating
factors dominate or whether negative-transfer rates are uniform across
architectures. For \textbf{RQ3}, since we are not aware of a CPL study
evaluating more than one model, cross-architecture consistency is
empirically unknown; we expect outcomes to be at least partially
architecture-mediated, given the fundamental inference-time differences
across BLAZE, COOBA, and TRANP-CNN, but make no
prediction about the degree of divergence or whether source quality metrics will
transfer across architectures.

\subsection{Benchmarked Models}
\label{sec:models}

A central design choice is evaluating CPL with three architecturally distinct
models rather than a single representative, guarding against conclusions
that are artifacts of one model's inductive bias. Each dominant paradigm is
represented by exactly one model: BLAZE (embedding-based), COOBA (GNN-based),
TRANP-CNN (CNN-based). Where the paper refers to ``embedding-based,''
``GNN-based,'' or ``CNN-based'' behavior, this is shorthand for the specific
model tested, not a claim about the paradigm generally; whether another
model of the same paradigm behaves the same way is untested here.

\paragraph{BLAZE \citep{chakraborty2025blaze}.}
A dual-encoder contrastive model~\citep{reimers2019sentencebert} that
retrieves the top-300 candidates via the shared FAISS index and re-ranks by
maximum cosine similarity. Its backbone is \texttt{BAAI/bge-code-v1}
\citep{bge_code}, frozen during training, with two modal-specific residual
adapter modules trained via in-batch NTXentLoss~\citep{oord2018cpc} and
hard-negative/hard-positive scaling, one for bug reports (prefixed
\texttt{search\_query:}), one for code chunks (prefixed
\texttt{search\_document:}). Each file is scored at evaluation time as the
maximum cosine similarity over its 512-token chunks (MaxSim), making the
architecture sensitive to the most relevant passage within a long file,
advantageous for large codebases.

\paragraph{COOBA \citep{zhu2020cooba}.}
A graph neural network encoding source files as AST graphs with GloVe-300
node features via graph convolutional layers (GCNConv) and global mean
pooling; bug reports are encoded by a bidirectional LSTM over GloVe-300
tokens, faithful to the original design. A shared feature extractor over
these text and AST representations is optimized to minimize ranking loss
while a Gradient Reversal Layer~\citep{ganin2016domain} before a binary
project discriminator prevents it from distinguishing source from target
representations. This adversarial objective is the core CPL mechanism.
COOBA reranks the shared FAISS top-300 pool.

\paragraph{TRANP-CNN \citep{huo2019deep}.}
A CNN-based reranker processing (bug report, candidate file) pairs through
parallel convolutional feature extractors (kernel sizes 3, 4, 5) into
scalar relevance scores, trained with margin ranking loss on FAISS top-300
candidates. Inputs are tokenized with the BGE tokenizer vocabulary but
mapped through embedding layers trained from scratch, following the
original design's learned-representation approach. An early deep
learning CPL approach, it provides a third architectural
perspective, convolutional text-code alignment, complementing BLAZE's
embedding-based and COOBA's graph-structural paradigms.

Table~\ref{tab:models} summarizes the three models' key properties,
together spanning the dominant implementation paradigms in current CPL
literature and providing a basis for testing whether CPL viability is a
property of the paradigm or of specific architectural choices.

\begin{table*}[t]
\centering
\caption{Summary of the three benchmarked bug localization models.}
\label{tab:models}
\begin{tabular}{@{}llll@{}}
\toprule
\textbf{Model} & \textbf{Architecture} & \textbf{CPL Mechanism} & \textbf{Retrieval} \\
\midrule
BLAZE \citep{chakraborty2025blaze} & Dual-encoder (BGE adapters) & Contrastive adapter fine-tuning & FAISS top-300 \\
COOBA \citep{zhu2020cooba}          & GNN on AST graphs          & Adversarial domain adaptation   & FAISS top-300 \\
TRANP-CNN \citep{huo2019deep}       & CNN reranker               & Shared representation learning  & FAISS top-300 \\
\bottomrule
\end{tabular}
\end{table*}

\paragraph{Adaptations and training configuration.}
\looseness=-1 BLAZE is adapted from its published design, with its BGE backbone
kept frozen and only its adapters trained (Table~\ref{tab:hyperparams}).
TRANP-CNN and COOBA predate the retrieval-augmented setup used
here and, with no original implementations available, were re-implemented
for this study, architecturally faithful to
their original publications (loss functions, encoders, adaptation
mechanisms preserved) with two deliberate departures: all three models
rerank the shared FAISS top-300 pool, and TRANP-CNN tokenizes with the BGE
vocabulary while training its embedding layers from scratch. Absolute
scores are therefore not directly comparable to the original papers'
(Section~\ref{sec:threats} discusses the construct implications).
Table~\ref{tab:hyperparams} lists each model's training configuration, held
fixed across all scenarios and pairs.

Throughout, ``architecture'' means the bundle distinguishing the three
models: input representation (BGE for BLAZE, GloVe-300 text/AST for COOBA,
learned tokens for TRANP-CNN), ranking mechanism (embedding similarity,
adversarially-adapted GNN score, CNN scalar score), and adaptation
mechanism (frozen backbone with adapters, domain-adversarial training,
standard fine-tuning). These dimensions co-vary with publication year
(BLAZE 2025, COOBA 2020, TRANP-CNN 2021)
and provenance (BLAZE adapted from its published design, the other two
re-implemented); Section~\ref{sec:discussion}
treats ``architecture'' as a bundle and attributes outcome differences to
it only as an interpretation, not a controlled causal claim, since these
dimensions are not independently manipulated here.

\begin{table*}[t]
\centering
\caption{Training configuration used for every experiment run. LR = learning rate;
effective batch = batch size $\times$ gradient-accumulation steps. Early stopping
monitors validation loss.}
\label{tab:hyperparams}
\small
\begin{tabular}{@{}llll@{}}
\toprule
\textbf{Setting} & \textbf{BLAZE} & \textbf{COOBA} & \textbf{TRANP-CNN} \\
\midrule
Optimizer       & AdamW & AdamW & AdamW \\
Learning rate   & $1\times10^{-4}$ & $1\times10^{-3}$ (discriminator $5\times10^{-4}$) & $1\times10^{-3}$ \\
Weight decay    & $10^{-2}$ & $10^{-5}$ & $10^{-5}$ \\
Effective batch & 8 (4 $\times$ 2) & 16 & 192 (96 $\times$ 2) \\
Max epochs      & 10 & 10 & 20 \\
Stopping        & Early stop, patience 3 & Fixed epoch budget & Early stop, patience 3 \\
Loss            & NTXent ($\tau = 0.3$) & Margin ranking $+$ adversarial ($\lambda_\mathrm{adv} = 0.1$) & Margin ranking \\
LR schedule     & Linear, 10\% warm-up & Constant & Constant \\
\bottomrule
\end{tabular}
\end{table*}

\subsection{Dataset Selection and Curation}
\label{sec:dataset}

\paragraph{Source Benchmarks.}
Projects were drawn from two established Python benchmarks: \textbf{SWE-bench}
\citep{jimenez2023swe}, contributing 11 of the 13 final projects with curated,
verified bug-report--to--commit-SHA ground truth across a broad domain
range, and \textbf{BeetleBox} \citep{chakraborty2025blaze}, contributing the
remaining 2 (ansible, localstack) to extend coverage to large-scale
infrastructure repositories absent from SWE-bench. Together they provide
the depth of verified ground truth needed for controlled CPL
evaluation~\citep{kalliamvakou2014promises}.

\paragraph{Selection Criteria.}
A project was retained if it had (i) \emph{bug report volume}: at least 100
unique bug reports with verified ground-truth file mappings, after
discarding reports with missing annotations, ambiguous multi-file patches
($>10$ files), or non-source patches. Two further conditions are properties
of the resulting \emph{set}, verified post hoc rather than filtered per
project: (ii) \emph{codebase size diversity} spanning at least one order of
magnitude, so size is a meaningful RQ2 moderator; and (iii) \emph{domain
coverage} including at least one project per target domain class, so no
single application type dominates. Both held for the set produced by
criterion (i) alone, so no project was added or removed to satisfy them.
This study restricts itself to Python to establish a language-homogeneous
baseline before cross-language extension (Section~\ref{sec:conclusion}).

\paragraph{Scope: 13 Python Projects.}
\looseness=-1 Applying criterion (i) yields \textbf{13 qualifying projects},
the \textbf{complete} qualifying set, none excluded after filtering. The
corpus spans meaningful variation: codebase size 35K--572K LoC (one order
of magnitude), four application domains, and bug corpus size 122--1,356
reports (Table~\ref{tab:dataset}).

The Python-only scope is also motivated by compute: 63 pairs $\times$ 4
scenarios $\times$ 3 models gives \textbf{756 pair-level results}, produced by
\textbf{606 training runs}, since WP-small depends only on the target and is
trained once per (model, target) with its result shared by every pair with
that target (Section~\ref{sec:scenarios}): per model, 63 pairs $\times$ 3
pair-dependent scenarios plus 13 WP-small runs. Each run involves training
from scratch, FAISS index construction, and per-bug evaluation,
approximately five to six months of wall-clock time in total
on NVIDIA L40S GPUs. Extending to Java would more than triple this budget,
so it is left to future work.

\begin{table*}[t]
\centering
\caption{The 13 Python projects used in this study. Size categories: Small ($<$100K LoC),
Medium (100--300K LoC), Large ($>$300K LoC). Source indicates the benchmark from which
each project's bug-report--commit ground truth was drawn. Recall@300 is the fraction
of bug reports for which the ground-truth file is retrieved within the shared
FAISS top-300 candidate pool when this project is the target; it bounds every
absolute metric for all three models, which rerank that same pool
(Section~\ref{sec:threats}).}
\label{tab:dataset}
\small
\begin{tabular}{@{}llrrllr@{}}
\toprule
\textbf{Project} & \textbf{Domain} & \textbf{LoC} & \textbf{Bug Reports} & \textbf{Size} & \textbf{Source} & \textbf{Recall@300} \\
\midrule
docker/compose          & Systems / Cloud     &  35,000 &  583 & Small  & SWE-bench & 99.8\% \\
jupyterlab/jupyterlab   & Data Science / ML   &  39,000 &  200 & Small  & SWE-bench & 47.0\% \\
lightning-ai/lightning  & Data Science / ML   &  51,000 &  374 & Small  & SWE-bench & 98.9\% \\
ipython/ipython         & Developer Tools     &  78,000 &  843 & Small  & SWE-bench & 74.6\% \\
prefecthq/prefect       & Data Science / ML   & 107,000 &  406 & Medium & SWE-bench & 69.7\% \\
mesonbuild/meson        & Developer Tools     & 121,000 &  945 & Medium & SWE-bench & 45.5\% \\
pydata/xarray           & Data Science / ML   & 142,000 &  122 & Medium & SWE-bench & 100.0\% \\
numpy/numpy             & Data Science / ML   & 277,000 &  803 & Medium & SWE-bench & 74.5\% \\
wagtail/wagtail         & Applications        & 286,000 &  407 & Medium & SWE-bench & 34.6\% \\
ansible/ansible         & Systems / Cloud     & 348,000 &  768 & Large  & BeetleBox & 31.3\% \\
scikit-learn/scikit-learn & Data Science / ML & 376,000 &  228 & Large  & SWE-bench & 62.3\% \\
qiskit/qiskit           & Applications        & 435,000 & 1,356 & Large  & SWE-bench & 56.1\% \\
localstack/localstack   & Systems / Cloud     & 572,000 &  472 & Large  & BeetleBox & 76.5\% \\
\bottomrule
\end{tabular}
\end{table*}

\paragraph{Domain Classification Protocol.}
To categorize each project's domain (Table~\ref{tab:dataset}), we followed a 
two-phase qualitative labeling process. In the first phase, the first author
 and a student volunteer examined the repositories' README files and 
 primary use cases to identify common themes, iteratively establishing 
 a taxonomy of four distinct domain classes. In the second phase, 
 both annotators independently classified each Python project 
 according to these established categories, agreeing on 11 of 13
 projects (84.6\% raw agreement). Disagreements primarily emerged
 around hybrid repositories—such as classifying 
 \texttt{jupyterlab/jupyterlab} as Data Science/ML 
 based on its intended workflow despite a frontend-heavy codebase, 
 or treating \texttt{ipython/ipython} as general Developer Tools. 
 All conflicting assignments were reviewed and resolved in a 
 single discussion round moderated by the second author to 
 reach a final consensus.

\paragraph{Pair Selection: 63 Directional Pairs.}
\looseness=-1 A na\"{i}ve all-pairs design over 13 projects yields 156 directed
pairs (1,872 runs), infeasible within our compute budget without
sacrificing statistical quality. We instead define a principled
\textbf{structured 63-pair set} in three groups (Table~\ref{tab:pairs}):
\textbf{Group A} (30 pairs), all bidirectional pairs among the six Data
Science/ML projects (jupyterlab, lightning-ai, prefect, xarray, numpy,
scikit-learn), the statistical backbone for controlled within-domain
comparison and commutativity analysis; \textbf{Group B} (14 pairs), both
directions between jupyterlab (39K LoC, second-smallest after docker/compose
at 35K) and each of the seven non-DS projects, isolating domain crossing
from target-size confound in the seven jupyterlab-as-target directions while
the reverse directions give complementary large-target observations; and
\textbf{Group C} (19 pairs), strategically selected cross-domain pairs
filling remaining gaps (large-to-large transfers, Systems/Applications
interactions, and ensuring every target has at least three candidate
sources for RQ3's source-selection analysis).

\begin{table*}[t]
\centering
\caption{Structure of the structured 63-pair set. Each pair is evaluated under
four training scenarios (Section~\ref{sec:scenarios}) and all three models
(Section~\ref{sec:models}), giving 252 pair-level results per model and 756 in total.}
\label{tab:pairs}
\small
\begin{tabular}{@{}llr@{}}
\toprule
\textbf{Group} & \textbf{Primary Purpose} & \textbf{Pairs} \\
\midrule
A, Within-Domain (DS$\times$DS)  & RQ1 statistical backbone; commutativity    & 30 \\
B, Cross-Domain Hub (jupyterlab) & Domain-crossing; target-size isolation      & 14 \\
C, Strategic Cross-Domain        & Large-target coverage; source selection     & 19 \\
\midrule
\textbf{Total}                      &                                              & \textbf{63} \\
\bottomrule
\end{tabular}
\end{table*}

\subsection{Data Preprocessing}
\label{sec:preprocessing}

The preprocessing pipeline runs once per project and persists outputs as
project-level databases, avoiding redundant recomputation across the
experiment runs.

\paragraph{Bug Report Curation.}
Raw bug reports are extracted from the two benchmarks and cross-referenced
against commit history for ground-truth file mappings, uniquely identified
by commit SHA; reports with missing SHAs, ambiguous patches ($>10$ files),
or non-source patches are discarded. The resulting bug metadata database
contains \texttt{\{bug\_id, bug\_text, ground\_truth\_files, commit\_sha\}},
where \texttt{ground\_truth\_files} is a list (Section~\ref{sec:metrics}
discusses the multi-file metric implications).

\paragraph{Source Code Embedding.}
Source files are extracted at each bug report's filing-time commit and
encoded with \textbf{BAAI/bge-code-v1} \citep{bge_code}, a
1536-dimensional Qwen2-based decoder pre-trained for code understanding;
files exceeding 512 tokens are split into overlapping chunks (512 tokens,
50-token overlap) and their embeddings averaged into a single file-level
\texttt{float32} vector, with bug report texts encoded analogously.
Producer-consumer multiprocessing (8 CPU workers) with batched GPU
inference (batch size 256) processes all 13 projects in approximately 6
hours on a single NVIDIA L40S GPU.

Per-project outputs are two binary databases, a \emph{blob embedding
database} (blob SHA $\to$ embedding) and a \emph{bug metadata database}
(bug ID $\to$ embedding and ground truth), from which all three models
draw identical first-stage retrieval candidates.

\subsection{Evaluation Pipeline}
\label{sec:pipeline}

\paragraph{Retriever-Reranker Pipeline.}
All three models share a two-stage pipeline. Retrieval builds a
\textbf{FAISS IndexFlatIP} index from L2-normalized BGE file embeddings
(inner product = cosine similarity) and queries the top-$K = 300$
candidates per bug report; recall at $K=300$ varies across projects
(Section~\ref{sec:threats}), and ground-truth files missed here are missed
at reranking. In reranking, TRANP-CNN and COOBA score (bug report,
candidate file) pairs directly, while BLAZE re-ranks the top-300 by maximum
cosine similarity between the fine-tuned bug report embedding and each
file's chunk embeddings (MaxSim). The raw FAISS ranking itself is weak
(mean MRR $= 0.029$; Table~\ref{tab:scenario_means}), low relative to
first-stage performance typically reported for dense retrieval on this
task; we attribute it to chunk-embedding averaging diluting long files'
signal, an attribution this paper does not test
(Section~\ref{sec:threats}). Its role is
candidate \emph{recall}; ranking precision is the rerankers' task.

\paragraph{Why Dense Retrieval, Not Lexical IR.}
The shared candidate-generation stage uses dense BGE embeddings rather than
a classical method like BM25~\citep{robertson2009probabilistic}, primarily
for representational consistency across the three rerankers rather than
lexical matching's vocabulary-mismatch limitation alone
(Section~\ref{sec:related}): BLAZE's own reranking already operates in this
embedding space (MaxSim over BGE chunks), so a shared dense backbone avoids
a second, uncontrolled representation mismatch on top of those discussed in
Section~\ref{sec:models}. This choice concerns candidate generation, not the cold-start
comparison: this paper does not report classical IR as a zero-label
comparator in its own right, so the cold-start viability result
(Section~\ref{sec:results}) is relative to WP-small, not the full set of
zero-label options, including lexical IR, open to an unlabelled
project.

\subsection{Training Scenarios}
\label{sec:scenarios}

\looseness=-1 The four training scenarios are the study's core methodological
contribution, addressing a specific evaluation gap: every prior CPL paper
compares against a within-project baseline trained on the \emph{majority}
of target labels, a ceiling data-scarce practitioners cannot reach, so
beating it leaves their question unanswered. Instead, the four scenarios form a
clean \textbf{ablation over the training-time supervision signal}, from
none, to limited target data only, to limited target data plus full source
data, to the ample-data ceiling, with the test set held constant throughout
(20\% of target bugs, seed 42, identical across models) so only training
composition changes, isolating the marginal value of each supervision
source.

\begin{itemize}
  \item \textbf{WP-small (WPS)} (Within-Project, Limited): trained on
    $\sim$20\% of the target corpus, drawn from the same 80\% pool as
    WP-large (nested; neither touches the 20\% test set), with the same
    target-side share as CP-transfer (below) so the only variable between
    the two scenarios is the presence of source data, ranging from 24
    labeled bugs (\texttt{pydata/xarray}) to 271 (\texttt{qiskit/qiskit}).
    This is the \emph{primary baseline} throughout, a deliberate departure from prior work's
    ample-data comparator: measuring CPL's benefit against a data-scarce
    baseline tests whether it delivers on its design intent. WP-small
    depends only on the target project, so it is run once per (model,
    target), 13 targets $\times$ 3 models, and that run serves every pair
    sharing the target. Like every other scenario, it is reported from a
    single run.

  \item \textbf{WP-large (WPL)} (Within-Project, Ample): trained on 80\% of the
    target corpus, the ceiling achievable with maximum within-project
    supervision. Not a CPL comparator but a reference: WP-large $-$
    CP-transfer quantifies what CPL still cannot achieve, while
    CP-transfer $-$ WP-small quantifies its gain over the practical
    baseline.

  \item \textbf{CP-cold-start (CPC)} (Cross-Project, Zero-Shot): trained on 100\%
    of source bugs with \emph{zero} target annotations, the minimum
    viable deployment for a project at day zero, testing whether a
    source-trained model can be applied with no annotation effort.

  \item \looseness=-1 \textbf{CP-transfer (CPT)} (Cross-Project, Fine-Tuned):
    jointly trained on 100\% of source bugs plus $\sim$20\% of target bugs
    (source and target batches in the same training step, not a sequential
    pretrain-then-fine-tune schedule), the same $\sim$20\% target-side
    share as WP-small.
\end{itemize}

Within each model's assigned training share, BLAZE and TRANP-CNN
additionally hold out 20\% for validation and early stopping
(Table~\ref{tab:hyperparams}'s ``Stopping'' row), drawn from the training
share itself and never from the 20\% test set; COOBA trains for a fixed
epoch budget with no validation split. The bug counts reported above for
WP-small (e.g.\ 24 for \texttt{pydata/xarray}) are the scenario's assigned
training-share size, not the smaller set actually used for weight updates
once this internal split is applied.

This framework applies identically to all three models, with no
model-specific adjustments, what makes the RQ3 cross-model comparison
valid and supports attributing effectiveness differences to the models
themselves rather than heterogeneous conditions. Table~\ref{tab:scenarios}
summarizes.

\begin{table*}[t]
\centering
\caption{Four evaluation scenarios with data configurations and roles. All share
the same held-out test set (20\% of target bugs, seed 42). WP-small, the primary
baseline, depends only on the target and is run once per (model, target), shared by
every pair with that target; the other scenarios are run once per pair.}
\label{tab:scenarios}
\small
\begin{tabular}{@{}llccp{6.0cm}@{}}
\toprule
\textbf{Scenario} & \textbf{Training Data} & \textbf{Target Budget} & \textbf{Runs} & \textbf{Role in Study} \\
\midrule
WP-small      & $\sim$20\% target bugs only         & Medium & 1 per target & Primary baseline; practical data-scarce setting, same target-side budget as CP-transfer \\
WP-large      & 80\% target bugs only               & High   & 1 per pair & Performance ceiling \\
CP-cold-start & 100\% source, 0\% target            & None   & 1 per pair & Zero-shot deployment viability \\
CP-transfer   & 100\% source + $\sim$20\% target    & Medium & 1 per pair & Core CPL scenario; same target-side budget as WP-small \\
\bottomrule
\end{tabular}
\end{table*}

\subsection{Evaluation Metrics}
\label{sec:metrics}

Five complementary metrics provide a complete picture of retrieval breadth and
ranking precision, computed over the re-ranked candidate list for each test bug
report and averaged over all test reports; a report whose ground truth is
not retrieved scores $0$ on every metric.

\textbf{Top-$K$ Recall} ($K \in \{1, 5, 10\}$): fraction of bug reports for which
at least one ground-truth file appears within the top $K$ positions. Top-1 captures
precision-at-one. Top-10 captures shortlist breadth.

A bug report's ground truth can span multiple files (patches touching up to 10
files are retained, Section~\ref{sec:dataset}), so MAP and MRR are computed under
the standard multi-relevant-document definitions rather than reducing to a shared
single-file formula.

\textbf{Mean Reciprocal Rank (MRR)}: for each bug report, the reciprocal rank of
the \emph{best-ranked} ground-truth file, $1/\min_j r_{i,j}$ over its ground-truth
files $j$, averaged across all bug reports $i$.

\textbf{Mean Average Precision (MAP)}: for each bug report with ground-truth files
at sorted ranks $r_{i,1} \le r_{i,2} \le \dots \le r_{i,k}$, the average precision
$\frac{1}{k}\sum_{m=1}^{k} m / r_{i,m}$, where $k$ is the number of ground-truth
files for the report and files absent from the reranked list contribute $0$ to
the sum, averaged across all bug reports. MAP and
MRR coincide only for the subset of bug reports with a single ground-truth file;
Table~\ref{tab:scenario_means} reports both because a substantial share of bug
reports in this corpus (45.6\%) have more than one.

\looseness=-1 MRR and MAP diverge from Top-$K$ when the correct file is retrieved but ranked
outside the top-$K$ shortlist. This divergence is itself informative. It reveals
whether failures are catastrophic (file not retrieved) or marginal (file ranked
too low).

\subsection{Domain Gap Measurement}
\label{sec:domaingap}

To quantify structural dissimilarity between source and target, a
\textbf{domain gap score} is computed for every pair by training a binary
logistic regression classifier (source files labeled 0, target labeled 1,
on file-level BGE embeddings) on a balanced 80/20 stratified split
($\ell_2$ regularization, default $C=1.0$, up to 1000 iterations); the
held-out test accuracy is the score. A score of 0.5 means the two projects
are indistinguishable in embedding space (no domain shift); 1.0 means fully
disjoint regions (maximum shift). This is grounded in
$\mathcal{H}$-divergence theory~\citep{bendavid2010theory}, which formalizes
domain separability as the accuracy of the best distinguishing classifier;
the linear probe here is a practical instantiation, avoiding a more
powerful classifier's risk of overfitting on small embedding datasets, and
has separately been widely used as a separability diagnostic in the
domain-adversarial literature~\citep{ganin2016domain}. Scores are computed
once per pair and shared across all three model evaluations.

Across the 63 pairs, scores are tightly concentrated at the high end
(min 0.976, median 0.989, max 0.997, mean 0.988, SD 0.006). Essentially
every pair is near-perfectly separable regardless of shared domain. A
logistic probe can separate almost any two distinct codebases this well,
so this restricted range should be kept in mind when interpreting the null
correlation between domain gap and CPL gain (Section~\ref{sec:results}): a
null result here is at least as consistent with too little variance to
detect an effect as with domain distance genuinely not mattering.

\subsection{Statistical Analysis Methods}
\label{sec:statmethods}

The analysis below follows recent guidance on statistical practice in empirical
software engineering, favoring non-parametric tests, explicit effect-size
reporting, and equivalence testing over reliance on unadjusted $p$-values
alone~\citep{oliveira2019evolution}.

\paragraph{RQ1, Wilcoxon Signed-Rank Test.}
\looseness=-1 To assess whether CP-transfer outperforms WP-small, a
\textbf{Wilcoxon signed-rank test}~\citep{wilcoxon1945,arcuri2011practical}
is applied per metric: observations are naturally paired (one CPT and one
WPS result per pair under identical conditions), and improvement
distributions cannot be assumed normal across so diverse a corpus. A
one-sided alternative is tested ($H_1$: CPT $>$ WPS). Effect sizes are
reported as the paired Cohen's $d_z$~\citep{cohen1988statistical,kampenes2007systematic,arcuri2014hitchhiker},
the mean of the per-pair CPT$-$WPS differences divided by their standard
deviation; because correlated pairs shrink this denominator relative to
the independent-samples case, $d_z$ is routinely larger than an
independent-samples $d$ computed on the same data, so the conventional
thresholds ($\geq 0.8$ large, $\geq 0.5$ medium, $\geq 0.2$ small) are
reported for comparability with prior literature as a convention rather
than a strict benchmark, alongside the matched-pairs
rank-biserial correlation $r_{rb}$, a signed-rank-aligned effect size that
is free of normality assumptions.

Significance markers denote unadjusted $p$-values ($\star\star\star$
$p<0.001$, $\star\star$ $p<0.01$, $\star$ $p<0.05$, n.s.\ $p\geq0.05$); with
five metrics tested per model family, the Bonferroni-adjusted threshold is
$\alpha/5 = 0.010$, and the text states explicitly which comparisons
survive it.

\paragraph{Equivalence Testing (TOST).}
Equivalence claims cannot rest on a non-significant difference test, since
absence of evidence is not evidence of absence. Where the paper reports
equivalence, we apply \textbf{two one-sided tests} (TOST)~\citep{schuirmann1987,lakens2017equivalence},
implemented as Wilcoxon tests against null hypotheses shifted by
$\pm\delta$, with margin $\delta = 0.05$ MRR. This margin is motivated, not
derived: at the viability threshold MRR $=0.20$ (ground truth around rank
five), 0.05 MRR is exactly the reciprocal-rank gap between rank 4 ($1/4$)
and rank 5 ($1/5$) for a single bug, defensible but not uniquely
correct, since MRR is a population average rather than a per-bug
statistic, so we report which conclusions are sensitive to it at
$\delta = 0.02$ and $0.10$ (Section~\ref{sec:results} notes where the
conclusion changes). Equivalence is declared when both one-sided tests
reject at $\alpha = 0.05$.

\paragraph{RQ2, Spearman Rank Correlation.}
\textbf{Spearman's $\rho$}~\citep{spearman1904} is computed between each
project-level feature and CPT$-$WPS gain, preferred over Pearson's $r$
since the corpus spans several orders of magnitude in codebase size and
bug count, making linear association inappropriate. These correlations
(eight features $\times$ three models) are exploratory and uncorrected for
multiple comparisons, unlike RQ1's Bonferroni-controlled family; only the
target-label-volume associations are strong enough to survive a Bonferroni
adjustment at this family size, and the text notes this where relevant.

\paragraph{RQ3, Source Selection Evaluation.}
\looseness=-1 \textbf{Kendall's $\tau$}~\citep{kendall1938new} is computed
between each heuristic's source ordering and the oracle ordering (by actual
CP-transfer MRR) for targets with at least three candidate sources.
\textbf{Hit@1} is the fraction of targets where the heuristic picks the
oracle-best source; mean \% of oracle performance achieved is also
reported.

\paragraph{Statistical Power and Sample Size.}
With $n=63$, a one-sided Wilcoxon test achieves approximately 98\% power to
detect a medium effect ($d=0.5$) at $\alpha=0.05$, adequately powered for
the effects observed for BLAZE ($d \approx 0.5$) and TRANP-CNN ($d \approx 0.95$),
and less so for an effect as small as COOBA's ($d \approx 0.27$). Power was
estimated via simulation under a normal approximation to the signed-rank
statistic, $10^4$ replications across a grid of effect sizes and sample
sizes.

\section{Results}
\label{sec:results}

\subsection{RQ1: Does CPL Outperform Within-Project Training Under Label Scarcity?}

\subsubsection{Scenario Performance Overview}

\looseness=-1 CP-transfer achieves MRR = 0.298 for BLAZE versus WP-small's 0.269 (both
at the same 20\% target-side budget; see
Section~\ref{sec:scenarios}), a mean gain of $+0.029$ (10.7\% relative). For
all three models the ordering WP-large $>$ CP-transfer $>$ WP-small $>$
CP-cold-start holds across all five metrics, with one exception (TRANP-CNN
Top-1, where CP-transfer edges WP-large, 0.134 vs.\ 0.132), and WP-small's
target-side data lifts it clearly above cold-start (BLAZE MRR 0.269 vs.\ 0.210;
MAP 0.232 vs.\ 0.181). For COOBA the gap between the middle two terms is small
(MRR 0.156 vs.\ 0.138). In
practical terms, a BLAZE CP-transfer team can expect to rank the correct file in
the top~5 in roughly 40\% of queries (Top-5 $= 0.395$) versus 35\% with a
within-project baseline of equal target-side budget, a modest but real
triage-workflow improvement.

For TRANP-CNN, CP-transfer achieves MRR = 0.214 versus WP-small's 0.165 (also
at the 20\% budget), a $+0.049$ gain (29.8\% relative), the largest of the
three models.
CP-transfer is \emph{statistically equivalent to WP-large} (0.219):
a TOST test bounds the difference within $\pm 0.05$
MRR ($p < 0.001$; two-sided Wilcoxon $p = 0.57$, 49.2\% win rate), holding
even at the tightest tested margin ($\delta = 0.02$, $p < 0.001$). TRANP-CNN
is the only model here for which 20\% target fine-tuning plus source data
reaches its own full within-project performance.

\looseness=-1 For COOBA, the mean gain is $+0.018$ MRR (12.9\% relative), and
cold-start collapses to near-zero (MRR = 0.027). Adding a full source corpus to
COOBA's 20\% target budget helps only modestly, and zero-shot transfer fails
outright without target-side signal for this architecture. Table~\ref{tab:scenario_means} details the
full per-scenario means.

\begin{table*}[t]
\centering
\caption{Mean bug localization performance across source$\to$target pairs per model and
scenario. WP-small is the primary baseline; CP-transfer is the primary CPL condition.
Bold denotes the best CPL scenario per model.
FAISS (retrieval stage) shows mean cosine-similarity ranking before any model reranking,
applicable to all three models. All values
on the same held-out test set (20\% of target bugs, seed 42).
MRR and MAP are averaged over \emph{all} held-out test reports (unretrieved ground
truth scores $0$).}
\label{tab:scenario_means}
\small
\begin{tabular}{@{}llrrrrr@{}}
\toprule
\textbf{Model} & \textbf{Scenario} & \textbf{MRR} & \textbf{MAP} & \textbf{Top-1} & \textbf{Top-5} & \textbf{Top-10} \\
\midrule
\multirow{5}{*}{BLAZE ($n=63$)}
  & FAISS retrieval & 0.029 & 0.023 & 0.003 & 0.019 & 0.072 \\
  & WP-small      & 0.269 & 0.232 & 0.185 & 0.352 & 0.423 \\
  & WP-large      & 0.344 & 0.302 & 0.258 & 0.450 & 0.508 \\
  & CP-cold-start & 0.210 & 0.181 & 0.126 & 0.295 & 0.378 \\
  & CP-transfer   & \textbf{0.298} & \textbf{0.257} & \textbf{0.213} & \textbf{0.395} & \textbf{0.468} \\
\midrule
\multirow{5}{*}{COOBA ($n=63$)}
  & FAISS retrieval & 0.029 & 0.023 & 0.003 & 0.019 & 0.072 \\
  & WP-small      & 0.138 & 0.119 & 0.061 & 0.218 & 0.333 \\
  & WP-large      & 0.192 & 0.165 & 0.112 & 0.280 & 0.370 \\
  & CP-cold-start & 0.027 & 0.023 & 0.008 & 0.035 & 0.059 \\
  & CP-transfer   & \textbf{0.156} & \textbf{0.134} & \textbf{0.075} & \textbf{0.237} & \textbf{0.339} \\
\midrule
\multirow{5}{*}{TRANP-CNN ($n=63$)}
  & FAISS retrieval & 0.029 & 0.023 & 0.003 & 0.019 & 0.072 \\
  & WP-small      & 0.165 & 0.140 & 0.084 & 0.247 & 0.323 \\
  & WP-large      & 0.219 & 0.192 & 0.132 & 0.314 & 0.387 \\
  & CP-cold-start & 0.038 & 0.033 & 0.010 & 0.055 & 0.095 \\
  & CP-transfer   & \textbf{0.214} & \textbf{0.185} & \textbf{0.134} & \textbf{0.306} & \textbf{0.370} \\
\bottomrule
\end{tabular}
\end{table*}

\subsubsection{Statistical Significance and Effect Size}

Table~\ref{tab:wilcoxon} presents Wilcoxon signed-rank test results and Cohen's $d$
for the CP-transfer versus WP-small comparison across all metrics and models.

For BLAZE, cross-project transfer significantly outperforms within-project
limited training across all five metrics ($p \leq 0.002$, surviving the
Bonferroni threshold on every metric). Win rates range from 60.3\% (MRR,
Top-1) to 69.8\% (Top-5), and effect sizes are small to medium by Cohen's
convention ($d = 0.43$--$0.60$; $r_{rb} = 0.43$--$0.62$).

For TRANP-CNN, the advantage is also significant across all five metrics and
likewise survives the Bonferroni threshold on \emph{every} metric (win rates
65.1\%--81.0\%; $d = 0.71$--$0.96$, $r_{rb} = 0.77$--$0.88$; all $p < 0.001$).
Its MRR/MAP effects ($d = 0.95$, $0.96$) are large and the largest of the three
models; their confidence intervals ($[0.75, 1.23]$ for MRR) do not overlap
BLAZE's ($[0.29, 0.69]$). TRANP-CNN's benefit also survives a cluster-aware
re-analysis by target project (Section~\ref{sec:threats}), which BLAZE's does
not. CP-transfer and WP-small are each estimated from a single run, so
individual-pair reproducibility should not be assumed from the aggregate
effect size.

\looseness=-1 For COOBA, the advantage is small: MRR and MAP are significant
at the unadjusted $0.05$ level ($p = 0.027$, $0.029$) but not at the
Bonferroni threshold, and no Top-$K$ metric is significant ($p = 0.07$--$0.13$).
Win rates (44.4\%--58.7\%) sit near chance and effect sizes are small
($d = 0.12$--$0.27$; $r_{rb} = 0.18$--$0.28$); the MRR and MAP confidence
intervals exclude zero, the Top-$K$ intervals do not.

We classify a model as exhibiting \emph{practically beneficial} CPL if
it meets both: significance on MRR/MAP at the Bonferroni threshold \emph{and}
at least a medium effect ($d \geq 0.5$, the conventional minimum for
consistent individual-deployment benefit). This is a post hoc criterion,
not a pre-registered threshold. TRANP-CNN meets both
criteria ($d = 0.95$--$0.96$ on MRR/MAP); BLAZE meets the significance
criterion and the effect-size criterion on MAP ($d = 0.53$) but falls just
short on MRR ($d = 0.49$); COOBA meets neither.

\looseness=-1 Taken together, the three models span a range of CPL benefit
rather than sharing one: a large effect for TRANP-CNN, a smaller one for
BLAZE, and a small, correction-sensitive one for COOBA (MRR $d$ =
$0.95$/$0.49$/$0.27$). All three point in the same direction, which suggests
the existence of a benefit is not an artifact of one model's inductive bias,
though with one model per paradigm this is not a methodological replication;
its size, however, depends strongly on the architecture.

\begin{table*}[t]
\centering
\caption{Wilcoxon signed-rank test results, Cohen's $d$ with 95\% CI, and matched-pairs
rank-biserial correlation $r_{rb}$ for CP-transfer vs.\ WP-small
(one-sided, $H_1$: CPT $>$ WPS). \textbf{Win \% here uses a strict zero threshold} (CPT
$>$ WPS MRR, any positive margin counts as a win); this is a different, more
permissive threshold than the $\pm 0.01$ MRR neutral band used for the
negative-transfer rate and for the within/cross-domain win rate in
Table~\ref{tab:crossdomain}, so the two should not be compared directly.
Stars denote unadjusted $p$-values:
$\star\star\star$ ($p < 0.001$), $\star\star$ ($p < 0.01$), $\star$ ($p < 0.05$), n.s.
The Bonferroni-adjusted threshold within each 5-metric family is $\alpha/5 = 0.010$;
all BLAZE and all TRANP-CNN metrics survive it, while no COOBA metric does
(COOBA MRR and MAP are significant only at the unadjusted level).
Note also that the Bonferroni family here
treats the five metrics as independent tests, though MRR and MAP are near
duplicates and the Top-$K$ values are nested, so this correction is
conservative.}
\label{tab:wilcoxon}
\small
\begin{tabular}{@{}llrrrrrrrr@{}}
\toprule
\textbf{Model} & \textbf{Metric} & \textbf{$n$} & \textbf{Win \%} & \textbf{$\Delta$ (mean)} & \textbf{$\Delta$ (median)} & \textbf{Cohen's $d$} & \textbf{95\% CI ($d$)} & \textbf{$r_{rb}$} & \textbf{Sig.} \\
\midrule
\multirow{5}{*}{BLAZE}
  & MRR    & 63 & 60.3\% & $+0.029$ & $+0.016$ & 0.49 & $[{+}0.29,{+}0.69]$ & 0.43 & $\star\star$ \\
  & MAP    & 63 & 66.7\% & $+0.025$ & $+0.013$ & 0.53 & $[{+}0.34,{+}0.74]$ & 0.50 & $\star\star\star$ \\
  & Top-1  & 63 & 60.3\% & $+0.028$ & $+0.021$ & 0.43 & $[{+}0.22,{+}0.66]$ & 0.44 & $\star\star$ \\
  & Top-5  & 63 & 69.8\% & $+0.044$ & $+0.026$ & 0.59 & $[{+}0.37,{+}0.83]$ & 0.59 & $\star\star\star$ \\
  & Top-10 & 63 & 66.7\% & $+0.045$ & $+0.019$ & 0.60 & $[{+}0.42,{+}0.82]$ & 0.62 & $\star\star\star$ \\
\midrule
\multirow{5}{*}{COOBA}
  & MRR    & 63 & 55.6\% & $+0.018$ & $+0.004$ & 0.27 & $[{+}0.04,{+}0.56]$ & 0.28 & $\star$ \\
  & MAP    & 63 & 58.7\% & $+0.015$ & $+0.005$ & 0.27 & $[{+}0.03,{+}0.53]$ & 0.27 & $\star$ \\
  & Top-1  & 63 & 44.4\% & $+0.014$ & $+0.000$ & 0.22 & $[{-}0.03,{+}0.47]$ & 0.23 & n.s. \\
  & Top-5  & 63 & 50.8\% & $+0.020$ & $+0.011$ & 0.23 & $[{-}0.01,{+}0.48]$ & 0.21 & n.s. \\
  & Top-10 & 63 & 50.8\% & $+0.007$ & $+0.006$ & 0.12 & $[{-}0.12,{+}0.38]$ & 0.18 & n.s. \\
\midrule
\multirow{5}{*}{TRANP-CNN}
  & MRR    & 63 & 81.0\% & $+0.049$ & $+0.045$ & 0.95 & $[{+}0.75,{+}1.23]$ & 0.84 & $\star\star\star$ \\
  & MAP    & 63 & 77.8\% & $+0.045$ & $+0.039$ & 0.96 & $[{+}0.76,{+}1.25]$ & 0.83 & $\star\star\star$ \\
  & Top-1  & 63 & 71.4\% & $+0.050$ & $+0.036$ & 0.76 & $[{+}0.58,{+}0.99]$ & 0.79 & $\star\star\star$ \\
  & Top-5  & 63 & 79.4\% & $+0.060$ & $+0.044$ & 0.86 & $[{+}0.68,{+}1.09]$ & 0.88 & $\star\star\star$ \\
  & Top-10 & 63 & 65.1\% & $+0.047$ & $+0.025$ & 0.71 & $[{+}0.54,{+}0.92]$ & 0.77 & $\star\star\star$ \\
\bottomrule
\end{tabular}
\vspace{2pt}

{\footnotesize Top-$K$ metrics are integer-valued and produce exact ties (equal Top-$K$ score) at
a rate the continuous MRR/MAP metrics do not: COOBA Top-1 has 11 tied pairs (28
wins, 24 losses), COOBA Top-5 has 7 (32 wins, 24 losses) and COOBA Top-10 has 9 (32
wins, 22 losses); Win \% counts ties as non-wins, so the COOBA Top-1 Win \% of
44.4\% sits below 50\% even though wins outnumber losses, and its median
$\Delta = +0.000$ reflects the ties.}
\end{table*}

\subsubsection{Win/Loss Asymmetry}

Whether CPL helps on average matters less than \emph{what happens when it
does not}: a model winning 60\% of the time but losing catastrophically in
the rest is worse than one winning 80\% with negligible losses. For BLAZE,
wins (median $+0.040$ MRR, $n = 38$) are about 2.3 times the size of losses
(median $-0.017$, worst $-0.045$, $n = 25$), but losses are frequent: 39.7\% of
pairs fall below WP-small under this strict (any-negative-gain) definition.
Because each scenario is a single run, individual small losses should not be
read as a precise gain-to-loss ratio.

\looseness=-1 For TRANP-CNN, wins clearly exceed losses (median $+0.057$, $n = 51$;
median $-0.010$, worst $-0.040$, $n = 12$), with the lowest loss exposure of the three models:
practitioners should expect roughly one in five deployments (19.0\%) to underperform
WP-small under the strict definition, with larger gains in the rest.

\begin{figure*}
\centering
\includegraphics[width=\textwidth]{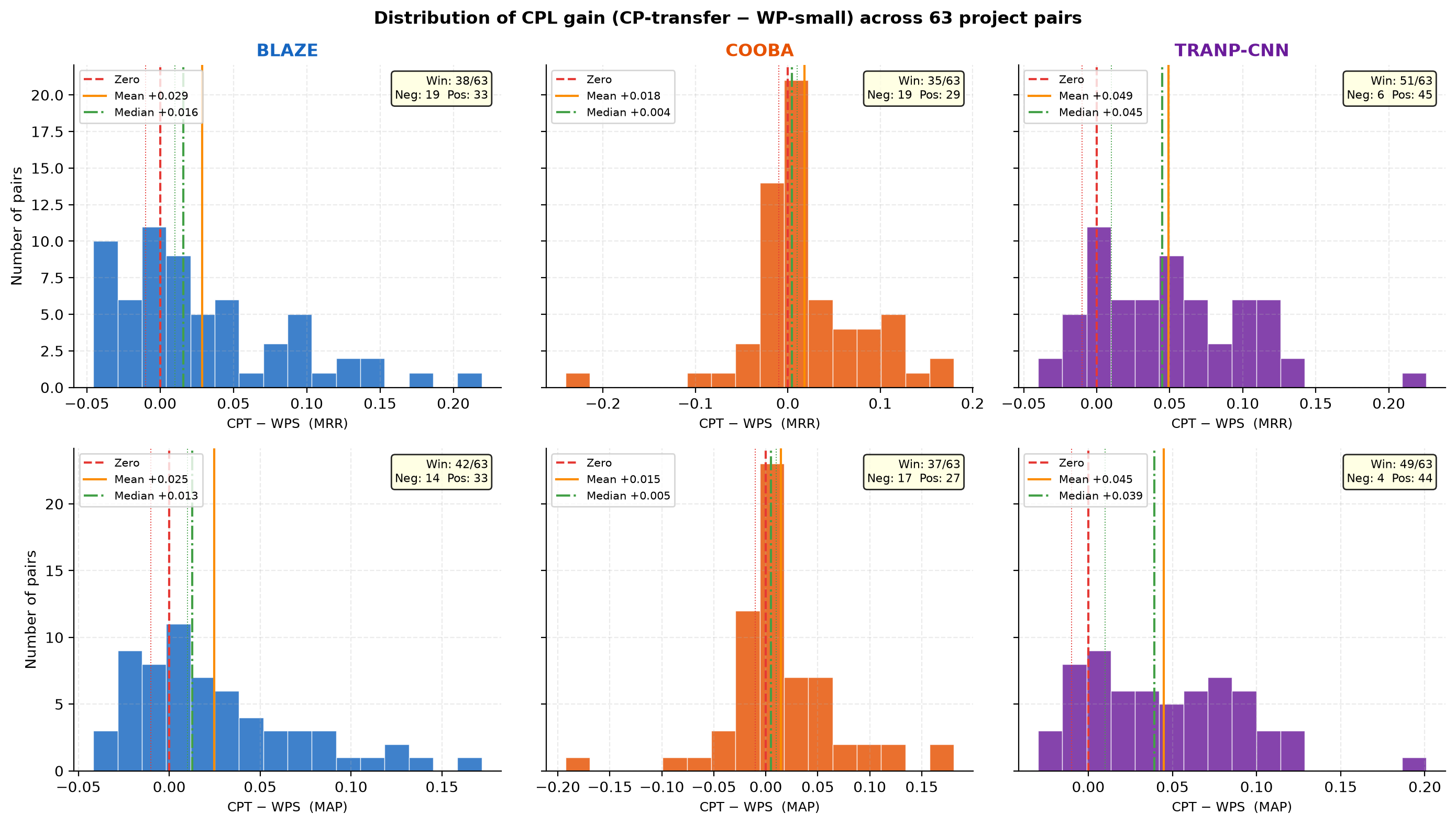}
\caption{Distribution of CP-transfer minus WP-small gain per source$\to$target pair
($n = 63$ pairs each, all three models), in MRR (top row) and MAP (bottom row);
the skew and SD values below refer to the MRR panels. Positive values indicate CPL benefit. BLAZE
(blue) shows a pronounced rightward skew (skew $+1.09$), TRANP-CNN (purple) a milder
rightward skew with the tightest spread ($+0.67$, SD $0.052$), and COOBA (orange) a
distribution centred near zero with a long \emph{left} tail (skew $-0.28$, driven by a
$-0.240$ MRR worst case) (see text for medians).
The vertical dashed line marks zero; dotted lines mark the $\pm$0.01 MRR threshold used
to classify positive/negative transfer.}
\label{fig:gain_distribution}
\end{figure*}

\looseness=-1 For COOBA, wins are slightly more frequent than losses (wins median
$+0.040$, $n = 35$; losses median $-0.019$, $n = 28$), and while the typical
loss is small, the tail is not: its worst case ($-0.240$ MRR) is five to six
times BLAZE's or TRANP-CNN's. Losses occur in 44.4\% of deployments, so a COOBA
practitioner is only marginally more likely to be helped than harmed. The
loss-exposure ordering (TRANP-CNN lowest at 19.0\%, then BLAZE at 39.7\%,
then COOBA at 44.4\%) mirrors the effect-size ordering.
Figure~\ref{fig:gain_distribution} illustrates the gain distributions.

\subsubsection{Zero-Label Deployment: The Cold-Start Finding}

Cold-start CPL (a source-trained model with zero target annotations) removes
the annotation bottleneck entirely, making its viability central to CPL's
utility case for data-scarce teams.

For BLAZE, CP-cold-start (MRR = 0.210) falls \emph{significantly short}
of WP-small (0.269; mean
difference $-0.060$ MRR, two-sided Wilcoxon $p < 0.001$). TOST equivalence
does not hold at the $\pm 0.05$ MRR margin ($p = 0.86$) or at the tighter $\delta = 0.02$ margin
($p = 0.999$); it is only recovered at a substantially looser
$\delta = 0.10$ margin ($p = 0.003$). The zero-label deployment claim for
BLAZE should therefore be read cautiously:
\emph{a BGE-based reranker deployed without
target labels retains about 78\% of the MRR of a within-project baseline
trained on 20\% of target labels}, not as equivalent performance. BLAZE re-scores
its own retrieval signal (with better MaxSim aggregation) rather than
applying an independent signal to a pool it never sees, unlike TRANP-CNN
and COOBA (Section~\ref{sec:threats} discusses this asymmetry); that
architectural distinction plausibly contributes to why BLAZE's cold-start gap to
WP-small is far smaller than TRANP-CNN's or COOBA's, even though it is not
statistically equivalent to it. This re-scoring asymmetry and the absence of
learned, target-specific scoring weights in BLAZE (Section~\ref{sec:architecture})
are two non-exclusive explanations for that smaller gap, which this study
cannot separate.

\begin{figure}
\centering
\includegraphics[width=\columnwidth]{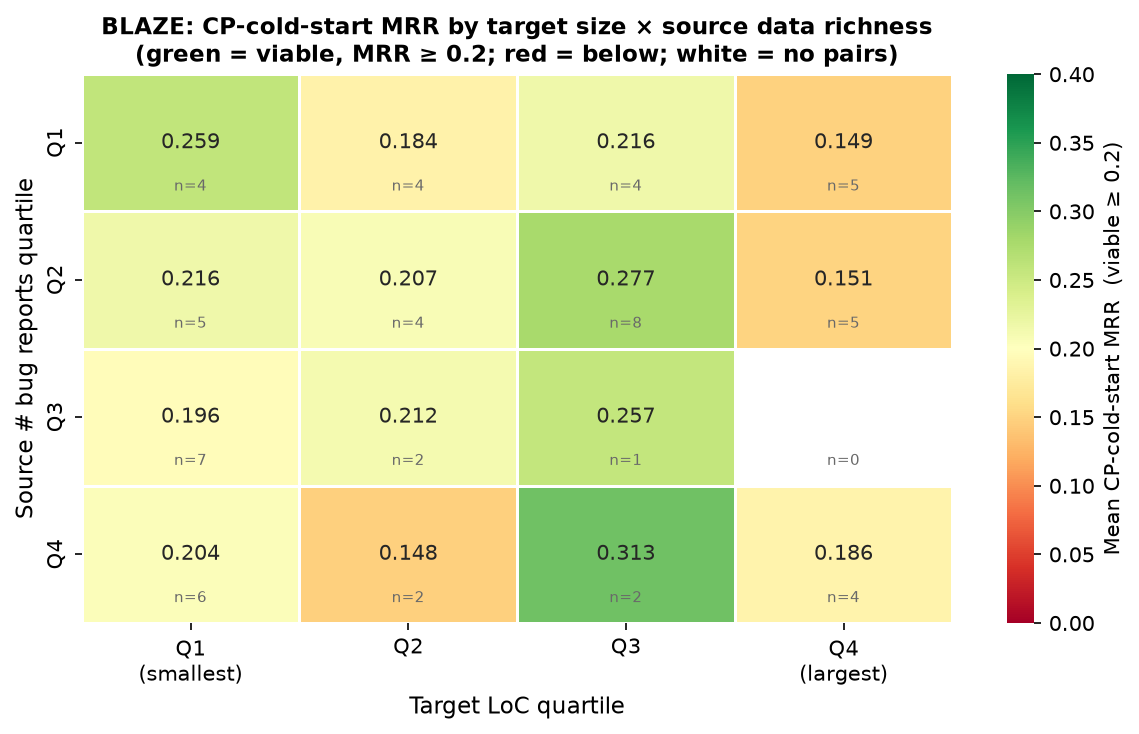}
\caption{BLAZE cold-start viability heatmap (CP-cold-start MRR), binned by target
LoC quartile (columns) and source bug-report-count quartile (rows), with cell
counts $n$ annotated. 9 of 15 populated cells exceed the MRR $= 0.20$ viability
threshold; three of the six exceptions are in the largest-target column
(source Q1/Q2/Q4 $\times$ target Q4, MRR $= 0.149/0.151/0.186$; source Q1/Q4
$\times$ target Q2, MRR $= 0.184/0.148$; source Q3 $\times$ target Q1,
MRR $= 0.196$, just below threshold). Green cells are at or above the threshold,
red-to-orange cells below it. One cell (source Q3 $\times$ target
Q4) is empty ($n=0$) and shown white. COOBA and TRANP-CNN are not shown: both
achieve MRR $<0.20$ almost everywhere in cold-start (0\% and 1.6\% of pairs
respectively).}
\label{fig:coldstart_heatmap}
\end{figure}

\looseness=-1 Figure~\ref{fig:coldstart_heatmap}'s cold-start viability heatmap shows
46.0\% of pairs (29/63) reaching MRR $\geq 0.20$, ground truth around rank
five, a triage-inspectable shortlist, so annotation effort can be
eliminated for roughly half of scenarios with a well-matched source, though
the majority case needs some target-side signal. Once a modest label set is
added (CP-transfer), the relevant question changes from viability to gain over
WP-small, a different metric: BLAZE beats WP-small in 63.3\% of within-domain
pairs (Table~\ref{tab:crossdomain}), reflecting its sensitivity to domain
alignment.

\looseness=-1 COOBA collapses to near-zero in cold-start (MRR = 0.027,
matching the raw FAISS baseline of 0.029, far below WP-small's 0.138,
$p < 0.001$, only 20\% of it). TRANP-CNN also fails, though less severely: 0.038 versus
WP-small's 0.165 ($p < 0.001$), 1.3$\times$ the FAISS baseline, some
residual utility from source-trained weights, but only 23\% of WP-small's
performance. Both scalar-ranking architectures need target-side calibration:
their scoring functions remain anchored to the source distribution without
it. The dichotomy is consistent with an architectural account: BLAZE's
embedding similarity carries over to unseen targets on average, while
TRANP-CNN's and COOBA's learned scores do not.

\begin{tcolorbox}[colback=gray!10, colframe=gray!50, title=\textbf{RQ1 Summary}]
CPL benefit under label scarcity differs in size across the models: large for
TRANP-CNN, smaller for BLAZE, small and correction-sensitive for COOBA.
\begin{itemize}[noitemsep,topsep=2pt]
  \item \textbf{BLAZE:} $d = 0.49$ (MRR), win rate 60.3\%, $p = 0.002$; cold-start
        \emph{not} equivalent to WP-small at the $\pm 0.05$ MRR margin
        (TOST $p = 0.86$; only equivalent at a looser $\pm 0.10$ margin,
        $p = 0.003$), viable in 46.0\% of pairs regardless; 30.2\% negative
        transfer.
  \item \textbf{TRANP-CNN:} $d = 0.95$ (MRR), win rate 81.0\%, $p < 0.001$;
        the largest effect, with a confidence interval clear of BLAZE's;
        CP-transfer equivalent to WP-large (TOST, $p < 0.001$); 9.5\%
        negative transfer, the lowest of the three.
  \item \textbf{COOBA:} $d = 0.27$ (MRR), significant only before correction
        ($p = 0.027$); cold-start fails; 30.2\% negative transfer.
\end{itemize}
\end{tcolorbox}

\subsection{RQ2: What Project-Level Characteristics Determine CPL Success or Failure?}

\subsubsection{Target Codebase Size}

\looseness=-1 Target codebase size moderates CPL benefit for BLAZE and COOBA,
and for TRANP-CNN non-monotonically. Table~\ref{tab:stratified} presents CPL
gain stratified by target size.

\looseness=-1 For BLAZE, gain is largest for small targets (mean $\Delta$MRR $=
+0.047$, $<$100K LoC), falls to $+0.027$ for medium targets, and vanishes for
large ones ($-0.001$): against a baseline with the same target-side
budget, BLAZE's CPL benefit is confined to small and medium codebases. For
TRANP-CNN, CP-transfer narrowly \emph{exceeds} WP-large for
small targets ($0.351$ vs.\ $0.345$) and for large targets ($0.093$ vs.\
$0.090$) (individual
pairs also cross this line; Section~\ref{sec:cases}); the gain is largest for
small targets ($+0.070$), smallest for medium ones ($+0.030$), and $+0.044$ for
large ones, so TRANP-CNN's gain is positive for every size group and not
monotone in LoC. For COOBA, the gain is concentrated in small targets
($+0.034$) and close to zero for medium ($+0.009$) and large ($+0.003$) ones,
so its modest aggregate benefit comes almost entirely from small codebases.

These gains warrant caution: FAISS recall at $K=300$ falls below 50\% for
four projects: ansible (31.3\%), wagtail (34.6\%), meson (45.5\%) and
jupyterlab (47.0\%), so part of the apparent LoC-driven degradation reflects a shared
retrieval ceiling rather than architecture alone. The within-model
comparison (CPT vs.\ WPS) is unaffected, since both face the same
constraint, but absolute gain magnitudes for large targets are suggestive
rather than definitive.

\begin{table*}[t]
\centering
\caption{CP-transfer vs.\ WP-small MRR gain by target codebase size group.
$\Delta$ = CP-transfer MRR $-$ WP-small MRR. WP-large MRR is shown for all models;
daggers mark the size-group aggregates where mean CP-transfer exceeds mean
WP-large.}
\label{tab:stratified}
\small
\begin{tabular}{@{}llrrrrr@{}}
\toprule
\textbf{Model} & \textbf{Size Group} & \textbf{$n$} & \textbf{WP-small MRR} & \textbf{WP-large MRR} & \textbf{CP-transfer MRR} & \textbf{$\Delta$ MRR} \\
\midrule
\multirow{3}{*}{BLAZE}
  & Small ($<$100K LoC)   & 25 & 0.254 & 0.348 & 0.301 & $+0.047$ \\
  & Medium (100K--300K)   & 24 & 0.296 & 0.376 & 0.323 & $+0.027$ \\
  & Large ($>$300K LoC)   & 14 & 0.252 & 0.284 & 0.250 & $-0.001$ \\
\midrule
\multirow{3}{*}{COOBA}
  & Small ($<$100K LoC)   & 25 & 0.209 & 0.294 & 0.243 & $+0.034$ \\
  & Medium (100K--300K)   & 24 & 0.107 & 0.152 & 0.116 & $+0.009$ \\
  & Large ($>$300K LoC)   & 14 & 0.065 & 0.080 & 0.068 & $+0.003$ \\
\midrule
\multirow{3}{*}{TRANP-CNN}
  & Small ($<$100K LoC)   & 25 & 0.280 & \textbf{0.345} & \textbf{0.351}$^{\dagger}$ & $+0.070$ \\
  & Medium (100K--300K)   & 24 & 0.112 & 0.162 & 0.142 & $+0.030$ \\
  & Large ($>$300K LoC)   & 14 & 0.049 & 0.090 & 0.093$^{\dagger}$ & $+0.044$ \\
\bottomrule
\multicolumn{7}{l}{\footnotesize $^{\dagger}$TRANP-CNN CP-transfer exceeds WP-large for small targets (0.351 vs.\ 0.345) and for large}\\
\multicolumn{7}{l}{\footnotesize targets (0.093 vs.\ 0.090). These are descriptive differences.}
\end{tabular}
\end{table*}

\begin{figure*}
\centering
\includegraphics[width=\textwidth]{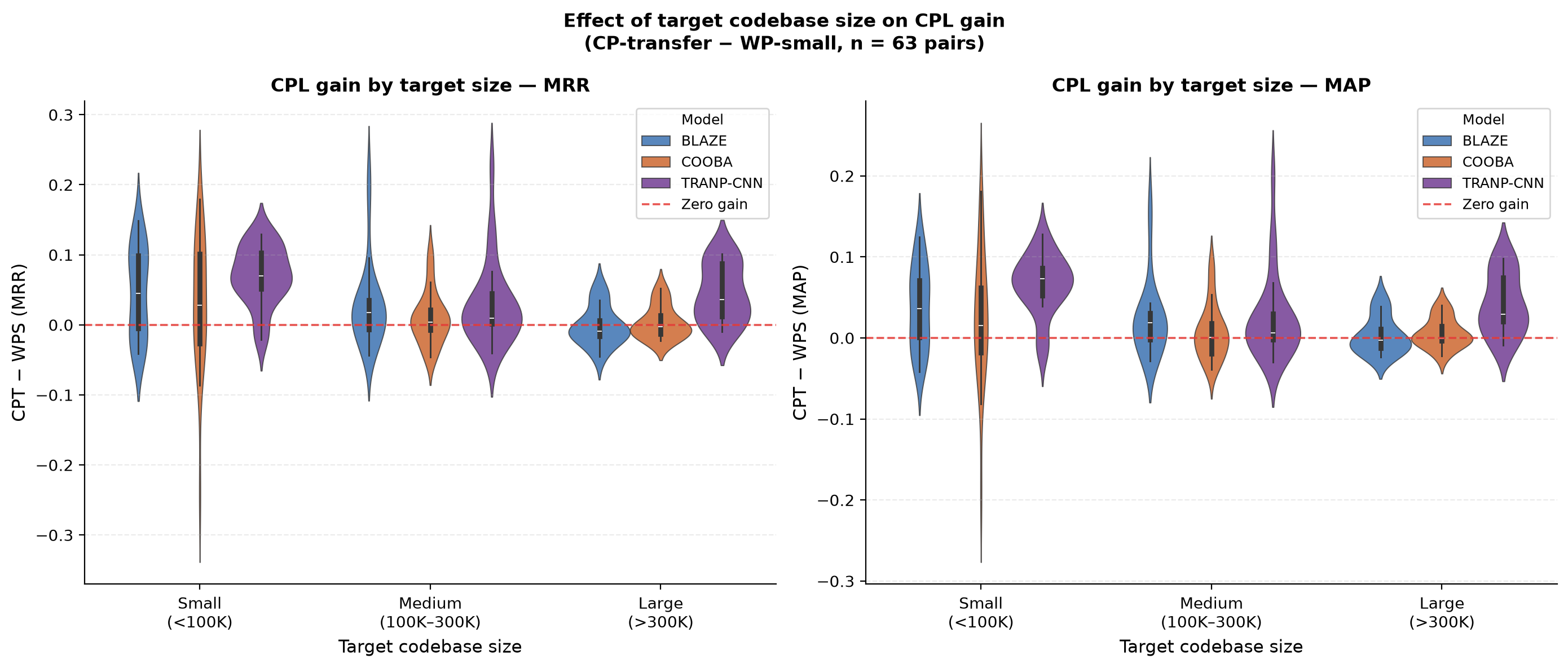}
\caption{Distribution of CP-transfer minus WP-small MRR gain by target codebase
size group (Small $<$100K, Medium 100K--300K, Large $>$300K LoC). Each violin shows
the spread of pair-level gains for a given model (color) and size group; see
Table~\ref{tab:stratified} for the corresponding means.}
\label{fig:gain_by_size}
\end{figure*}

Figure~\ref{fig:gain_by_size} shows the distribution of CPL gain by target size group.
For BLAZE, the small and medium groups are centered positively, while the large
group is centered at zero. For COOBA, only the small group is shifted positively.
For TRANP-CNN, all groups are positive, with the small group exceeding the WP-large ceiling.

\subsubsection{Negative Transfer}

Negative transfer, where source-project knowledge actively harms target
performance, is CPL's central risk, determining whether it is a safe
default or requires careful source selection. We define it as CP-transfer
MRR $<$ WP-small MRR by more than 0.01 ($|\Delta| \leq 0.01$ treated as
no-effect). This $\pm 0.01$ band is a sensitivity threshold, not a
noise-calibrated one (Section~\ref{sec:threats}).

For TRANP-CNN, 6 of 63 pairs (9.5\%, roughly one in ten) are negative, the
lowest rate of the three, worst at ansible $\to$ prefect ($-0.040$), numpy
$\to$ prefect ($-0.038$), and wagtail $\to$ docker/compose ($-0.021$),
concentrated where high-specificity codebases are targets (prefect accounts
for 3 of the 6, docker/compose for 2). For BLAZE, 19 of 63 (30.2\%) are negative,
worst at jupyterlab $\to$ localstack ($-0.045$), numpy $\to$ prefect ($-0.044$),
and ansible $\to$ lightning ($-0.041$); prefect is again the most
frequently harmed target (4 of 19). For COOBA, 19 of 63 (30.2\%,
nearly a third) are negative, worst at wagtail $\to$ docker/compose
($-0.240$, the single largest loss anywhere in the study), ipython $\to$
docker/compose ($-0.086$), and numpy $\to$ lightning ($-0.072$); jupyterlab is
its most frequently harmed target (3 of 19). Practitioners should treat
CPL for all three models as requiring empirical validation per deployment, not a
universally safe upgrade. TRANP-CNN's rate is about a third of the other two
models', which are equal; negative transfer is common for BLAZE and COOBA and
present, though rarer, for TRANP-CNN. Because each scenario is reported from a
single run, individual pair labels near the $\pm 0.01$ band are less certain
than the aggregate rates, and the recurrence of particular targets among the
losses should not be read as evidence of a specific mechanism, which we did not
test.

\begin{figure*}
\centering
\includegraphics[width=\textwidth]{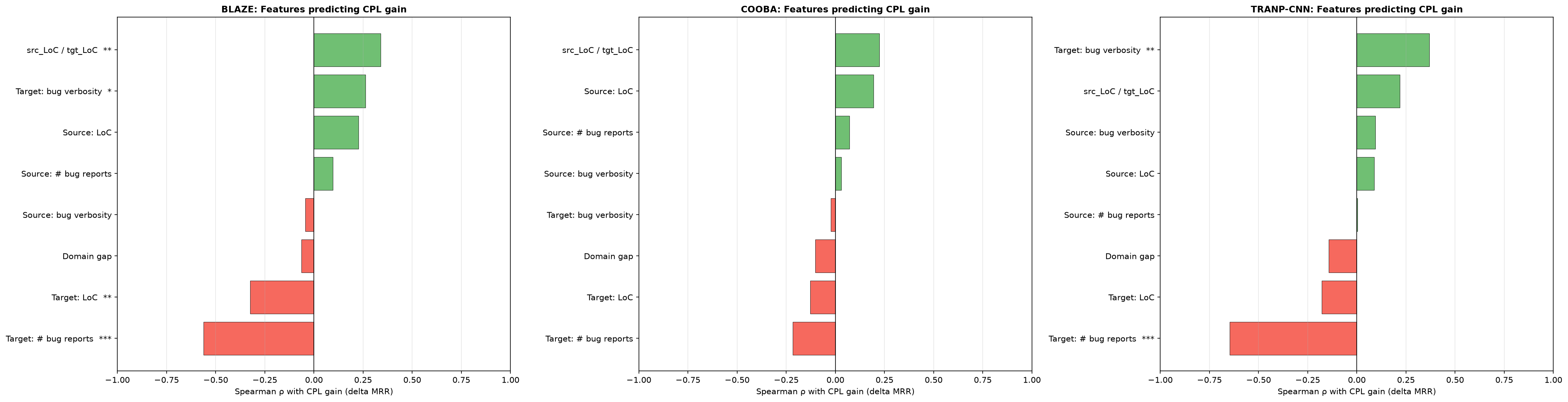}
\caption{Spearman correlations between project-level metadata features (target
and source LoC, target and source bug-report count, LoC ratio, domain gap
accuracy, target and source bug-report verbosity) and CP-transfer minus
WP-small MRR gain, shown for all three models (BLAZE, COOBA, TRANP-CNN); see text
for which correlations reach significance.}
\label{fig:feature_corr}
\end{figure*}

Figure~\ref{fig:feature_corr} plots Spearman correlations between project-level
metadata features and CPL gain. \emph{Target label volume}, which sets the
absolute size of the WP-small training set (Section~\ref{sec:scenarios}), is
the strongest predictor we find, and a negative one: for TRANP-CNN, target
bug-report count correlates with CPL gain at $\rho = -0.65$ ($p < 0.001$), and
for BLAZE at $\rho = -0.56$ ($p < 0.001$). The fewer labeled bugs a
target already has, the more it stands to gain from CPL, a pattern
Case~3 (Section~\ref{sec:cases}) illustrates on a single label-poor target.
These are the only two correlations that survive a Bonferroni correction
over all 24 tests. Target LoC, the source$\to$target LoC ratio, and target
bug-report verbosity are also nominally significant for BLAZE ($\rho = -0.32$,
$\rho = 0.34$, and $\rho = 0.26$; $p = 0.010$, $0.007$ and $0.037$), and
verbosity for TRANP-CNN ($\rho = 0.37$, $p = 0.003$), though verbosity is itself
moderately correlated with target label volume ($\rho = -0.50$), so it is a
related facet of the same under-resourced-target pattern rather than an
independent signal. Similarity-based
features that are not proxies for label volume are not predictive:
domain gap accuracy is non-significant for all three models (BLAZE
$\rho = -0.06$, $p = 0.62$; COOBA $\rho = -0.10$, $p = 0.42$; TRANP-CNN
$\rho = -0.14$, $p = 0.26$), and for COOBA \emph{no} measured feature reaches
significance. The largest are the LoC ratio ($\rho = 0.22$, $p = 0.077$) and
target label volume ($\rho = -0.22$, $p = 0.088$). COOBA's outcomes are
therefore not predictable from the project-level features measured here,
consistent with its small systematic gain; for BLAZE and TRANP-CNN, target
label volume is a useful, if imperfect, guide.

\subsubsection{Source-Target Domain Alignment}

\begin{figure*}
\centering
\includegraphics[width=\textwidth]{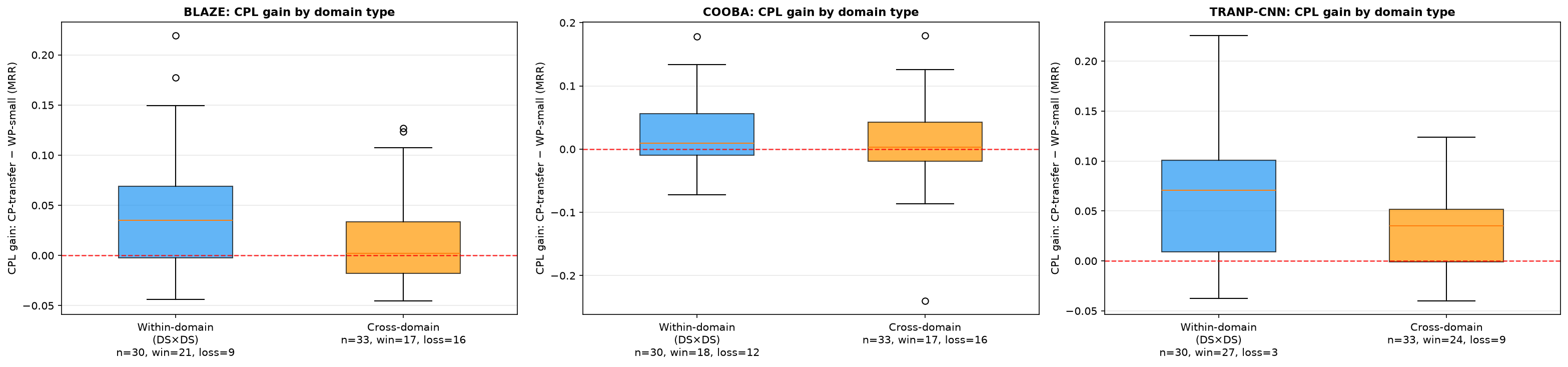}
\caption{CPL gain distributions (CP-transfer $-$ WP-small MRR) for within-domain
(DS$\times$DS) versus cross-domain pairs across all three models. The win/loss counts annotated
on each box use a strict zero threshold (CPT $>$ WPS), whereas
Table~\ref{tab:crossdomain} reports win rates under the $\pm 0.01$ MRR
neutral band, so the two are not directly comparable.}
\label{fig:cross_domain}
\end{figure*}

\looseness=-1 Table~\ref{tab:crossdomain} breaks CPL gain down by whether source and
target share a domain (within-domain, Group A) or not (cross-domain, Groups
B/C); Figure~\ref{fig:cross_domain} shows the corresponding distributions.
All within-domain pairs are Data Science/ML (Group A), so this contrast also
reflects domain identity, not just alignment.

Domain alignment matters most for BLAZE. For BLAZE, within-domain pairs gain
substantially more ($+0.043$ MRR, 63.3\% win rate) than cross-domain
($+0.016$, 42.4\%), a 20.9pp drop, the clearest domain sensitivity among the
three models; cross-domain negative transfer is not rare (13 of 33). For COOBA
the contrast is smaller (within-domain 50.0\% vs.\ cross-domain 42.4\%, 7.6pp),
and for TRANP-CNN smaller still (73.3\% vs.\ 69.7\%, 3.6pp), although
TRANP-CNN's mean gain is about twice as large within-domain ($+0.066$ vs.\
$+0.034$). Same-domain sourcing is therefore a useful heuristic chiefly for
BLAZE; for the other two models it shifts win rates only modestly in this
corpus.

Group B (14 of 33 cross-domain pairs) is built entirely around jupyterlab
as the pivot; excluding it, cross-domain win rates become 26.3\% (BLAZE),
42.1\% (COOBA), 63.2\% (TRANP-CNN). BLAZE's cross-domain
performance excluding jupyterlab is markedly weaker than its
jupyterlab-inclusive figure, so its domain sensitivity is not an artifact of
the pivot, while TRANP-CNN's drops by 6.5pp and COOBA's barely moves,
underscoring that jupyterlab's role as a pivot project affects each model
very differently.

\looseness=-1 The domain gap metric (logistic regression accuracy on file
embeddings, Section~\ref{sec:domaingap}) shows no significant correlation
with gain for any of the three models (BLAZE $\rho = -0.06$, $p = 0.62$;
COOBA $\rho = -0.10$, $p = 0.42$; TRANP-CNN $\rho = -0.14$, $p = 0.26$).
The domain binary is a stronger signal than this continuous metric for BLAZE
(a 20.9pp win-rate gap), while for TRANP-CNN and COOBA neither signal is
strongly informative.

\subsubsection{Commutativity: Does Transfer Direction Matter?}

For 28 symmetric pairs ($A \to B$ and $B \to A$ both evaluated), we test
whether the smaller-LoC target consistently gains more, as the LoC-stratified
results would predict. TRANP-CNN shows this in 53.6\% of pairs
(15/28, one-sided binomial $p = 0.43$ against a 50\% baseline) and BLAZE in
60.7\% (17/28, $p = 0.17$), neither distinguishable from chance at the $0.05$
level, so LoC asymmetry is not the dominant factor for either
model; other factors, not tested here, must account for the rest.
COOBA is highest at 75.0\% (21/28, $p = 0.006$), the only model for which
the smaller target gains more often than chance, consistent with its gain
being concentrated in small targets (Table~\ref{tab:stratified}); with 28
symmetric pairs this is a suggestive pattern rather than an established one.

\begin{tcolorbox}[colback=gray!10, colframe=gray!50, title=\textbf{RQ2 Summary}]
CPL benefit is largest for small-codebase targets; within-domain sourcing
helps BLAZE most:
\begin{itemize}[noitemsep,topsep=2pt]
  \item \textbf{BLAZE:} 30.2\% negative transfer, 20.9pp
        within-domain advantage (the largest of the three), 42.4\% cross-domain win rate.
  \item \textbf{TRANP-CNN:} 9.5\% negative transfer (lowest), little domain
        sensitivity in win rate (3.6pp), CPT
        exceeds WPL for small and large targets.
  \item \textbf{COOBA:} 30.2\% negative transfer; gain concentrated in small
        targets ($+0.034$), near zero elsewhere.
\end{itemize}
Target label volume best predicts gain for TRANP-CNN and BLAZE
($\rho = -0.65$, $-0.56$); no measured feature predicts COOBA's outcomes.
\end{tcolorbox}

\begin{table*}
\centering
\caption{CPL gain by domain alignment (within-domain = same application category;
cross-domain = different category). $\Delta$MRR = mean CP-transfer $-$ WP-small MRR.
Win Rate uses the $\pm 0.01$ MRR neutral band, a different (stricter) threshold
than Table~\ref{tab:wilcoxon}'s zero-cutoff Win \% (see that table's caption).
BLAZE shows a clear domain sensitivity (20.9 percentage-point win-rate drop);
for COOBA (7.6pp) and TRANP-CNN (3.6pp) the within-domain advantage is smaller.}
\label{tab:crossdomain}
\small
\begin{tabular}{@{}llrrr@{}}
\toprule
\textbf{Model} & \textbf{Domain Type} & \textbf{$n$} & \textbf{$\Delta$MRR (mean)} & \textbf{Win Rate} \\
\midrule
\multirow{2}{*}{BLAZE}
  & Within-domain (DS$\times$DS) & 30 & $+0.043$ & 63.3\% \\
  & Cross-domain                 & 33 & $+0.016$ & 42.4\% \\
\midrule
\multirow{2}{*}{COOBA}
  & Within-domain (DS$\times$DS) & 30 & $+0.025$ & 50.0\% \\
  & Cross-domain                 & 33 & $+0.011$ & 42.4\% \\
\midrule
\multirow{2}{*}{TRANP-CNN}
  & Within-domain (DS$\times$DS) & 30 & $+0.066$ & 73.3\% \\
  & Cross-domain                 & 33 & $+0.034$ & 69.7\% \\
\bottomrule
\end{tabular}
\end{table*}

\subsection{RQ3: How Consistent Are CPL Outcomes Across Model Implementations?}

\subsubsection{Cross-Model Direction Agreement}

\looseness=-1 Cross-model agreement tests whether the CPL benefit observed for one
architecture predicts benefit for another on the same pair: if CPL were a
property of the pair, all models should agree; if architecture-mediated,
they should diverge. Table~\ref{tab:model_agreement} reports pairwise
direction agreement across all three model combinations.

\begin{table}[t]
\centering
\caption{Pairwise cross-model direction agreement on CPL gain (CPT$-$WPS MRR, $n = 63$).
Agreement = fraction of pairs where both models show the same directional outcome
(positive, neutral, or negative gain). Spearman $\rho$ measures the continuous
correlation of pair-level CPL gains. $\star\star\star$: $p < 0.001$; $\star$: $p < 0.05$.}
\label{tab:model_agreement}
\small
\begin{tabular}{@{}lrr@{}}
\toprule
\textbf{Model Pair} & \textbf{Agreement} & \textbf{Spearman $\rho$} \\
\midrule
BLAZE vs.\ COOBA      & 44\% & $0.240$ (n.s.) \\
BLAZE vs.\ TRANP-CNN  & 54\% & $0.419^{\star\star\star}$ \\
COOBA vs.\ TRANP-CNN  & 54\% & $0.286^{\star}$ \\
\bottomrule
\end{tabular}
\end{table}

\begin{figure*}
\centering
\includegraphics[width=\textwidth]{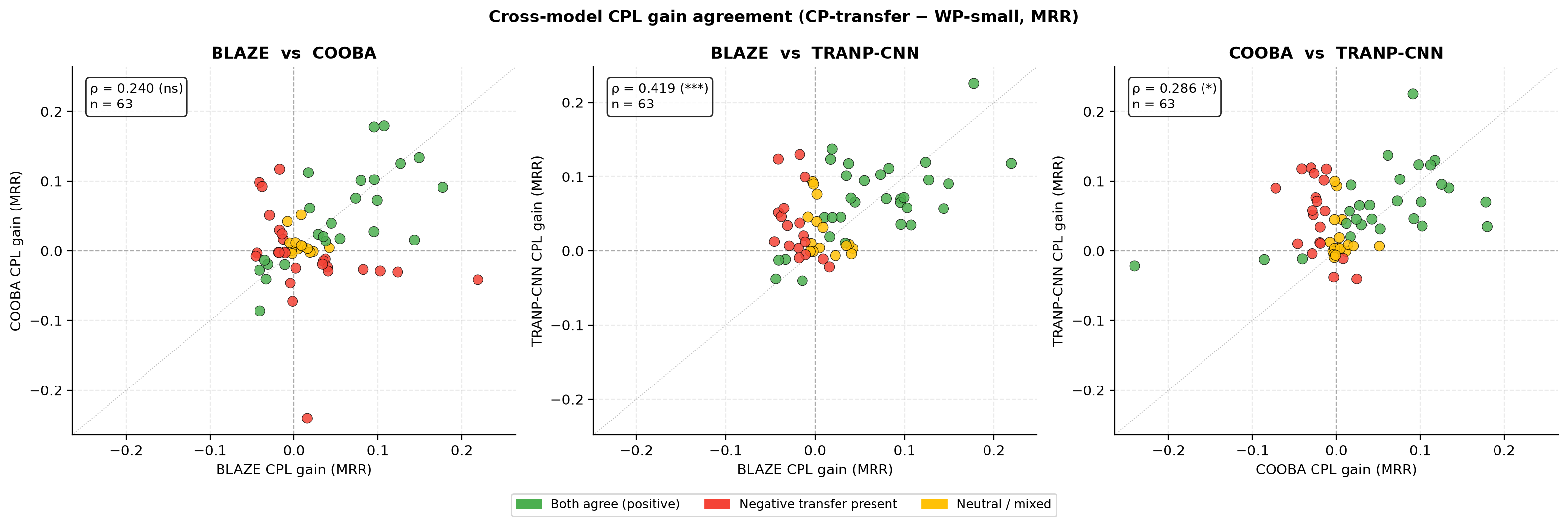}
\caption{Cross-model scatter of CPT$-$WPS MRR gain across 63 shared
source$\to$target pairs, for all three model pairs: BLAZE vs.\ COOBA (left),
BLAZE vs.\ TRANP-CNN (center), and COOBA vs.\ TRANP-CNN (right); see
Table~\ref{tab:model_agreement} for the corresponding correlations.}
\label{fig:agreement_scatter}
\end{figure*}

BLAZE and TRANP-CNN agree in 54\% of pairs with a moderate positive
correlation ($\rho = 0.419$, $p < 0.001$), the strongest pair-level
relationship of the three; COOBA agrees with TRANP-CNN in 54\% of
pairs ($\rho = 0.286$, $p = 0.023$) and with BLAZE in only 44\%
($\rho = 0.240$, $p = 0.058$). COOBA's pair-level outcomes are thus only weakly
related to either of the other models'. The most pronounced divergences:
BLAZE gains $+0.219$ MRR on lightning$\to$xarray while COOBA loses $-0.041$,
and COOBA loses $-0.240$ on wagtail$\to$docker/compose (its single worst
outcome across all 63 pairs) while BLAZE is mildly positive ($+0.016$).
Figure~\ref{fig:agreement_scatter} shows the cross-model scatter plots.

\looseness=-1 This has a natural interpretation: BLAZE and TRANP-CNN both
assign relevance scores at inference time (cosine similarity vs.\ CNN
scalar) with cross-project knowledge injected at training time, while
COOBA's adversarial mechanism trains a discriminator to erase
source-target distinctions, a fundamentally different adaptation
strategy producing different sensitivity to the same pairs. CPL outcomes
therefore vary substantially across architectures on the same pair, which
Section~\ref{sec:architecture} interprets via the model's adaptation
mechanism: a practitioner should not assume one model's outcome predicts
another's, and the 44\%--54\% agreement range is consistent with
architecture being at least as strong a determinant of outcomes as the
pair itself.

\subsubsection{Source Quality as a Transfer Predictor}

\looseness=-1 A natural source-selection heuristic is to pick the source that
performs best within-project. We test this via Spearman $\rho$ between the
source's WP-large MRR and its achieved CP-transfer MRR. For BLAZE,
$\rho = -0.02$ ($p = 0.89$): no predictive power in this corpus, so
source selection need not be preceded by within-project evaluation for BLAZE.
For TRANP-CNN, $\rho = -0.36$ ($p = 0.004$): a significant \emph{negative}
predictor, sources on which TRANP-CNN performs best within-project transfer
worse in absolute CP-transfer MRR (though not in gain over WP-small,
$\rho = -0.03$, $p = 0.83$). For COOBA, $\rho = -0.295$
($p = 0.019$): likewise a significant \emph{negative} predictor, sources on which
COOBA performs best within-project actually transfer worse, conjectured
(not verified by inspecting AST structure) to reflect overfitting to
source-specific patterns that become a liability on the target. As for
TRANP-CNN, the association does not extend to gain over WP-small
($\rho = -0.14$, $p = 0.28$). COOBA practitioners should therefore avoid
within-project source performance as a selection criterion entirely.

All three models thus agree: source selection based on within-project
quality is unreliable or counter-productive across all three paradigms.

\subsubsection{Source Selection Heuristics}

Table~\ref{tab:source_selection} evaluates two source-selection heuristics
against the oracle, across targets with at least three candidate sources:
\emph{most-bugs} (largest bug corpus) and \emph{domain-gap} (lowest
embedding-space gap). Hit@1 is how often a heuristic picks the oracle-best
source, Kendall $\tau$ the quality of its full ranking, and \% oracle the
achieved CP-transfer MRR relative to the best possible source.

\looseness=-1 For BLAZE, most-bugs gets Hit@1 $= 23.1\%$, $\tau = -0.050$
(near-zero); domain-gap is worse on Hit@1 ($15.4\%$) but weakly positive on
$\tau$ ($+0.020$). Neither ranks the full list reliably, yet BLAZE reaches
89.3\%--92.2\% of oracle with either. Source selection has low stakes
here; even a poorly chosen source is near-oracle. For TRANP-CNN, most-bugs
gets Hit@1 $=30.8\%$, $\tau = +0.112$, the largest positive $\tau$ among the
evaluated heuristics; domain-gap gets Hit@1 $=23.1\%$,
$\tau = -0.080$; oracle recovery is 91.1\%--92.6\% either way, though the
9.5\% negative-transfer risk still makes holdout validation
advisable. For COOBA, most-bugs gets Hit@1 $=38.5\%$, $\tau = -0.195$;
domain-gap gets Hit@1 $=23.1\%$, $\tau \approx 0$ ($-0.007$, expected: domain
similarity carries no signal when the CPL mechanism itself erases
domain-specific features); oracle recovery is uneven, 73.8\% for
most-bugs, 84.8\% for domain-gap, the widest heuristic gap and, at its
low end, the lowest oracle recovery in this study.

\begin{figure*}
\centering
\includegraphics[width=0.32\textwidth]{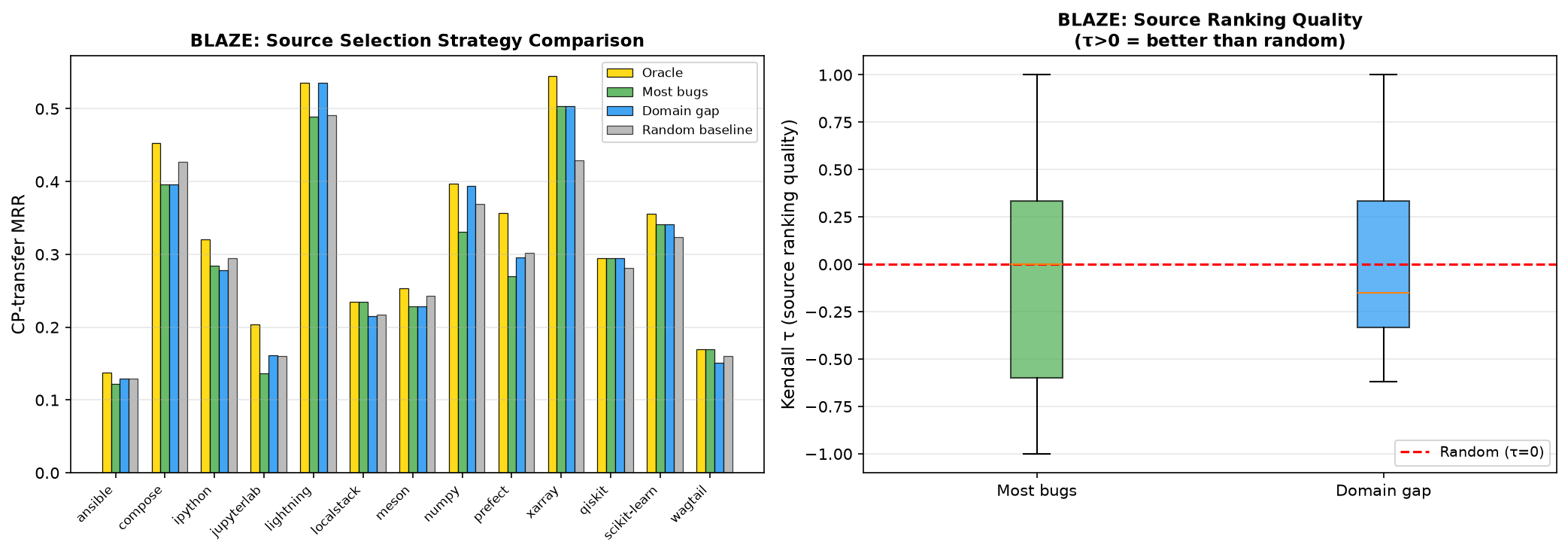}\hfill
\includegraphics[width=0.32\textwidth]{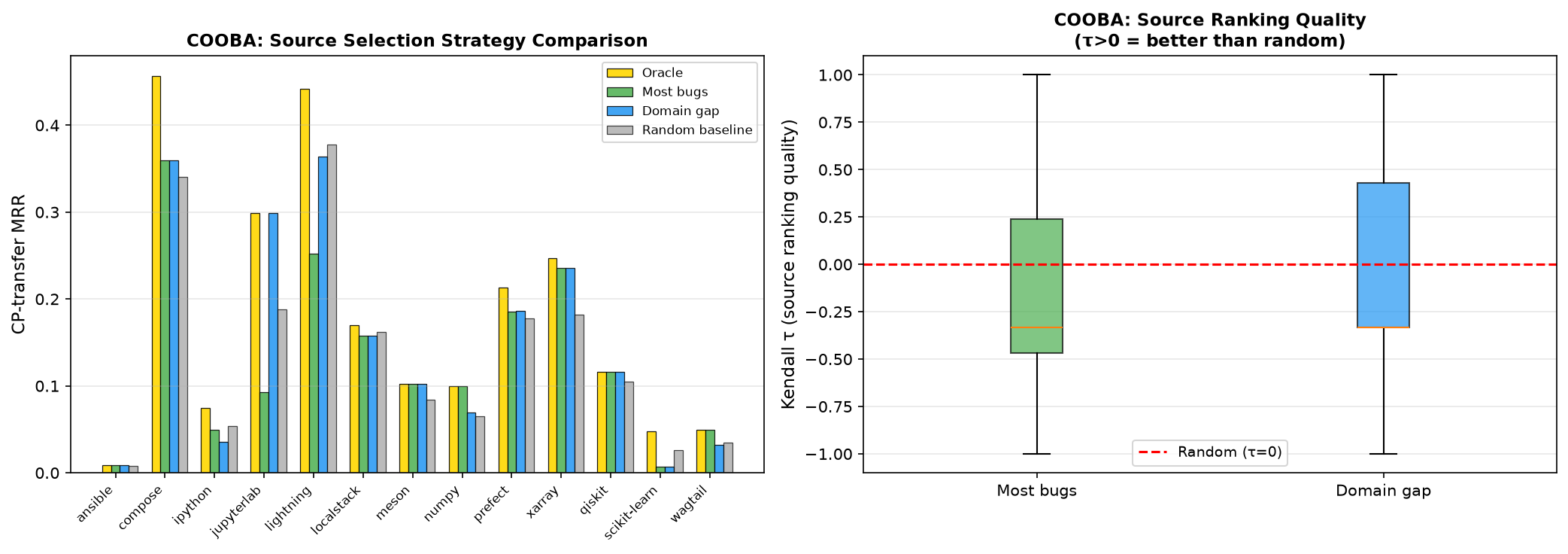}\hfill
\includegraphics[width=0.32\textwidth]{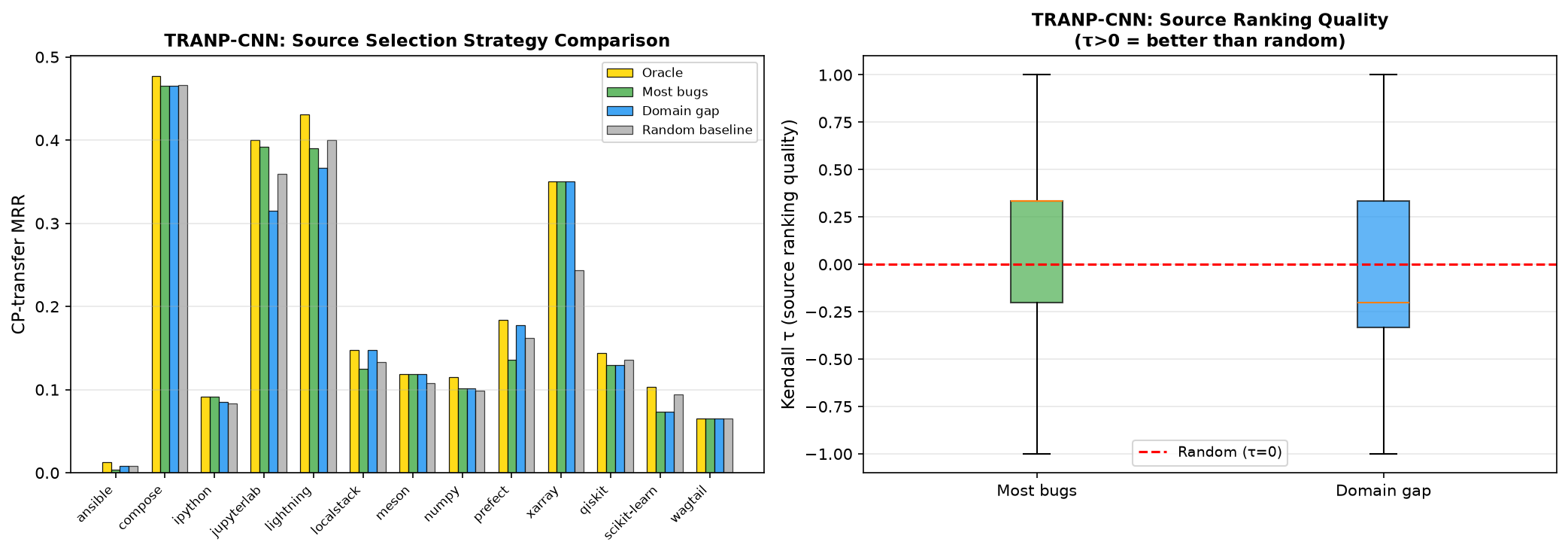}
\caption{Source selection quality for BLAZE (left pair of panels), COOBA (center), and TRANP-CNN (right). In each
pair, the bar chart gives CP-transfer MRR per target under the oracle source, the most-bugs and
domain-gap heuristics, and the random-selection baseline; the boxplot gives the distribution across
targets of Kendall $\tau$ for the two heuristics ($\tau > 0$ is better than random).
Table~\ref{tab:source_selection} reports the corresponding means.}
\label{fig:ss_quality}
\end{figure*}

The takeaway is consistent across heuristics and models
(Figure~\ref{fig:ss_quality} visualizes per-target oracle, heuristic and random MRR and the $\tau$ distributions): no
heuristic reliably ranks sources ($\tau < 0.2$ throughout), yet all three
models reach 74--93\% of oracle with any of them. A random-selection
baseline (mean CP-transfer MRR across all candidate sources, computed from
the same per-target data) recovers a comparable 78--90\% (Table~\ref{tab:source_selection}),
confirming that this reflects source choice being genuinely low-stakes for
BLAZE and TRANP-CNN rather than the two heuristics specifically being
effective. COOBA's lower absolute
recovery makes selection matter more: domain-gap recovery (84.8\%) exceeds
random selection (77.6\%) while most-bugs (73.8\%) does not, though the
analysis does not establish statistical superiority. Empirical holdout validation on a small target sample
remains the safest strategy.

\begin{table*}[t]
\centering
\caption{Source selection heuristic evaluation. Two heuristics tested: \emph{most-bugs}
(source with largest bug corpus) and \emph{domain-gap} (source with lowest embedding-space
domain gap), plus a \emph{random selection} baseline (mean CP-transfer MRR across all
candidate sources). ``\% oracle'' = mean CP-transfer MRR of selected source as a fraction of
oracle CP-transfer MRR, computed as a ratio of means across targets (summed
selected-source MRR over summed oracle MRR). No heuristic reliably ranks sources ($\tau < 0.2$ in all cases),
yet oracle recovery remains high (74--93\%) for all strategies, comparable to random
selection (78--90\%), so this is evidence that source choice is low-stakes rather than
that these two heuristics are specifically effective. $n = 13$
targets with at least three candidate sources; Hit@1 differences of one or
two targets are not distinguishable at this sample size.}
\label{tab:source_selection}
\small
\begin{tabular}{@{}llrrr@{}}
\toprule
\textbf{Model} & \textbf{Strategy} & \textbf{Hit@1} & \textbf{Kendall $\tau$} & \textbf{\% Oracle} \\
\midrule
\multirow{4}{*}{BLAZE}
  & Oracle (upper bound)   & 100\%  & N/A    & 100\%  \\
  & Most-bugs heuristic    &  23.1\% & $-0.050$ & 89.3\% \\
  & Domain-gap heuristic   &  15.4\% & $+0.020$ & 92.2\% \\
  & Random selection       &  N/A    & N/A    & 89.9\% \\
\midrule
\multirow{4}{*}{COOBA}
  & Oracle (upper bound)   & 100\%  & N/A    & 100\%  \\
  & Most-bugs heuristic    &  38.5\% & $-0.195$ & 73.8\% \\
  & Domain-gap heuristic   &  23.1\% & $-0.007$ & 84.8\% \\
  & Random selection       &  N/A    & N/A    & 77.6\% \\
\midrule
\multirow{4}{*}{TRANP-CNN}
  & Oracle (upper bound)   & 100\%  & N/A    & 100\%  \\
  & Most-bugs heuristic    &  30.8\% & $+0.112$ & 92.6\% \\
  & Domain-gap heuristic   &  23.1\% & $-0.080$ & 91.1\% \\
  & Random selection       &  N/A    & N/A    & 89.3\% \\
\bottomrule
\end{tabular}
\end{table*}

\begin{tcolorbox}[colback=gray!10, colframe=gray!50, title=\textbf{RQ3 Summary}]
Cross-model agreement ranges from 44\% (BLAZE vs.\ COOBA) to 54\% (BLAZE
vs.\ TRANP-CNN, COOBA vs.\ TRANP-CNN); BLAZE's and TRANP-CNN's gains are the most
strongly correlated ($\rho = 0.419$, $p < 0.001$), COOBA's only weakly with
either. Source within-project quality does not predict
transfer for BLAZE, and is a significant \emph{negative} predictor
for TRANP-CNN ($\rho = -0.36$) and COOBA ($\rho = -0.295$). All three models stay near-oracle under source
selection (74--93\% of oracle with any source), making heuristic source
ranking low-priority relative to empirical holdout validation.
\end{tcolorbox}

\subsection{Illustrative Cases}
\label{sec:cases}

The aggregate findings above describe CPL behavior across 63 pairs and three
models. Three project pairs make these patterns concrete: cold-start
viability, architecture divergence on the same pair, and CP-transfer
exceeding the ample-data ceiling on a label-poor target.

\paragraph{Case 1: BLAZE cold-start viability (lightning-ai\,$\to$\,xarray).}
lightning-ai/lightning (51K LoC) sources pydata/xarray
(142K LoC), a within-domain pair. BLAZE cold-start (CPC, MRR $= 0.384$)
exceeds WPS ($0.326$) by $+0.058$, the source-trained model beats 20\%
annotated target-only training with zero target labels. TRANP-CNN and COOBA
cold-start both collapse ($0.009$ and $0.028$), far below their own WPS.

\begin{center}
\small
\begin{tabular}{lcccc}
\toprule
Model & WPS & WPL & CPC & CPT \\ \midrule
BLAZE     & 0.326 & 0.551 & \textbf{0.384} & 0.545 \\
TRANP-CNN & 0.125 & 0.236 & 0.009 & 0.243 \\
COOBA     & 0.144 & 0.182 & 0.028 & 0.103 \\ \bottomrule
\end{tabular}
\end{center}

\noindent For a team beginning xarray localization with no labeled data, a
pre-trained BLAZE model sourced from lightning outperforms annotating 20\% of
local bugs. Neither other architecture replicates this, for the mechanism
reasons given in Section~\ref{sec:architecture}.

\paragraph{Case 2: Architecture divergence on the same pair (wagtail\,$\to$\,docker/compose).} 
wagtail/wagtail (286K LoC) sources docker/compose (35K LoC),
a cross-domain pair. BLAZE gains mildly ($+0.016$ MRR, CPT$-$WPS); COOBA experiences
\emph{negative transfer} ($-0.240$, the largest single loss in the study) on
the identical target; TRANP-CNN is
slightly negative ($-0.021$), already strong within-project.

\begin{center}
\small
\begin{tabular}{lcccc}
\toprule
Model     & WPS   & WPL   & CPC   & CPT \\ \midrule
BLAZE     & 0.437 & 0.495 & 0.246 & \textbf{0.453} \\
COOBA     & 0.445 & 0.450 & 0.084 & 0.205 \\
TRANP-CNN & 0.478 & 0.452 & 0.049 & 0.457 \\ \bottomrule
\end{tabular}
\end{center}
\noindent The same source simultaneously produces a modest BLAZE gain and
COOBA's strong negative transfer on the identical target: a COOBA practitioner would
call wagtail harmful, a BLAZE practitioner mildly helpful. COOBA's loss here
is several times larger than any other model's on any pair, so it is unlikely
to be a matter of run-to-run variation alone, though both scenarios are single
runs (Section~\ref{sec:threats}). No model-agnostic viability judgment is
possible here.

\paragraph{Case 3: CP-transfer exceeds full within-project training (prefect\,$\to$\,jupyterlab).} 
\textit{prefecthq/prefect} (107K LoC) sources
\textit{jupyterlab/jupyterlab} (39K LoC), a within-domain pair whose target has
the second-smallest bug corpus in the study (200 mappings, so WP-small is
$\sim$40 labeled bugs). TRANP-CNN's CP-transfer ($0.371$) exceeds WP-large
($0.287$) by $+0.084$: joint training on the full source corpus plus 20\%
target labels beats four times as much target-only data. BLAZE does the same
more modestly ($0.203$ vs.\ $0.164$, $+0.039$), while COOBA is level with its
own ceiling ($0.253$ vs.\ $0.257$).

\begin{center}
\small
\begin{tabular}{lcccc}
\toprule
Model & WPS & WPL & CPC & CPT \\ \midrule
BLAZE     & 0.054 & 0.164 & 0.135 & 0.203 \\
COOBA     & 0.119 & 0.257 & 0.005 & 0.253 \\
TRANP-CNN & 0.280 & 0.287 & 0.100 & \textbf{0.371} \\ \bottomrule
\end{tabular}
\end{center}

\noindent Two of the three architectures clear their own ample-data ceiling on
this pair, so the effect is not unique to one model's inductive bias, but the
margins differ by a factor of two and COOBA does not clear it at all. The
mechanism is visible in BLAZE's row: its WPS ($0.054$) is the weakest
within-project result of any model on this target, yet its CP-transfer
($0.203$) is nearly four times that. When a target's own labeled history is thin
enough, a full source corpus supplies more usable signal than the target can
supply about itself. This is why CPL is most valuable early in a project's
life, before a meaningful labeled history has accumulated, and it is the same
relationship the target-label-volume correlation captures across all 63 pairs
(Section~\ref{sec:results}).

\section{Discussion}
\label{sec:discussion}

\subsection{Practical Implications}

The results support six concrete recommendations for practitioners considering
CPL deployment.

\paragraph{1. For BLAZE, CPL showed a modest pair-level benefit with small losses in this corpus.}
CP-transfer outperformed WP-small in most deployments (60.3\%,
Table~\ref{tab:wilcoxon}), with losses small (median $-0.017$ MRR, worst
$-0.045$; Section~\ref{sec:results}), though 30.2\% of pairs show negative
transfer and the benefit is not significant once pairs are clustered by
target (Section~\ref{sec:threats}). \emph{Zero-label deployment retains about
78\% of WP-small's MRR on average, and is viable in under half of pairs.}
Deployment can begin without target annotations, with labels accumulated
subsequently for fine-tuning.

\paragraph{2. Target codebase size moderates benefit; validate empirically for large targets.}
CPL gain is largest for small-codebase targets and drops for medium targets;
for BLAZE it is essentially zero for large targets
(Table~\ref{tab:stratified}), while TRANP-CNN retains a gain there and COOBA's
small gain is confined to small targets. Practitioners with large-codebase targets and
BLAZE-style tools should validate CPL empirically before relying on it.

\paragraph{3. Choose sources by domain, not by ranking heuristics or within-project quality.}
No tested heuristic reliably ranks source projects for any model, yet all
three achieve near-oracle CP-transfer performance regardless of source choice
(Table~\ref{tab:source_selection}), so source selection has low stakes for
BLAZE and TRANP-CNN. Domain category membership is the most actionable
criterion, though the within-domain evidence here comes entirely from Data
Science pairs: same-domain sources give a markedly higher win rate for
BLAZE (63.3\% vs.\ 42.4\%) and only modestly higher ones for the other two
models, and for BLAZE any conveniently available within-domain source beats
investing in ranking by corpus size or embedding proximity. Within-project
quality should not be used as a selection criterion for any architecture;
for TRANP-CNN and COOBA it is a significant \emph{negative} predictor (Section~\ref{sec:results}),
conjectured to reflect overfitting to source-specific patterns, and
COOBA's lowest oracle recovery and joint-highest negative-transfer rate make
holdout validation advisable there.

\paragraph{4. With 20\% target labels, TRANP-CNN reaches full within-project training performance.}
With 20\% of bugs labeled, TRANP-CNN's
CP-transfer performance, averaged over all 63 pairs, is statistically equivalent to WP-large, and exceeds
it for small-codebase targets (Section~\ref{sec:results}), the highest CPL
ceiling \emph{relative to its own baseline} of the three models, though BLAZE
still leads on absolute MRR and Top-K throughout
(Table~\ref{tab:scenario_means}). This carries some
negative-transfer risk (Section~\ref{sec:results}), though in this corpus it is
the lowest of the three models and only mildly concentrated in cross-domain pairs
(win rate 73.3\% within-domain vs.\ 69.7\% cross-domain), so empirical validation
remains the safer guard.

\paragraph{5. Empirical holdout validation is the safest guard against negative transfer.}
Target label volume predicts CPL gain for BLAZE and TRANP-CNN, but no
measured feature reliably predicts pair-level gain for COOBA
(Section~\ref{sec:results}). When deployment risk is high, the safest
approach is a small annotated target holdout before committing to
cross-project transfer, especially for BLAZE and COOBA (each with roughly one in
three deployments negative), and to a lesser extent TRANP-CNN (roughly one in
ten); a project able to hold one out is, by definition, not fully in the
zero-label regime this study is written for (Section~\ref{sec:threats}
gives a rough usable size).

\paragraph{6. CP-transfer's benefit has a compute cost that WP-small does not.}
CP-transfer trains on the full source corpus plus the same target-side share
as WP-small, so its cost scales with source-corpus size rather than
staying fixed (Section~\ref{sec:threats}). BLAZE's zero-label cold-start
deployment avoids this entirely, part of its appeal in
recommendation 1; for TRANP-CNN, which needs the full CP-transfer run to
reach its ceiling, this cost is worth weighing against WP-small alone when
the source corpus is large and the expected gain modest.

\subsection{Architectural Interpretation of CPL Profiles}
\label{sec:architecture}

CPL outcomes vary substantially across architectures on the same
source$\to$target pairs, and we offer a mechanistic account: models that
rank at inference time via cosine similarity carry cross-project semantics
in their representations, while models applying learned scalar scores need
target-side calibration and cannot transfer without it. The three models
span distinct paradigms of how a bug localization model evaluates
candidates at inference time, under the identical protocol established in
Section~\ref{sec:design}. Each model consumes its
native input representation (BGE for BLAZE, learned tokens for TRANP-CNN,
GloVe-300 text/AST for COOBA), treated as a constituent of the architecture
under test. We interpret each model's \textbf{evaluation mechanism}, what
its learned parameters and representations do to rank candidates at test
time, as the dimension most consistent with the divergent profiles below,
though as Section~\ref{sec:threats} notes, representation quality is bundled into
each architecture rather than independently controlled, so ``evaluation
mechanism'' should be read as our best-supported interpretation rather than
an isolated, independently-verified cause.

\looseness=-1 The divergent CPL profiles are consistent with that mechanism.

\paragraph{BLAZE (embedding-based dual-encoder).}
BLAZE scores candidates at inference time using cosine similarity between
frozen BGE backbone embeddings, with no learned scalar weight applied at
ranking time. The ordering is determined by the frozen pre-trained
representations and the trained adapters, with no learned scalar at
ranking time; the frozen backbone, built on a diverse multilingual code
corpus, already carries cross-project semantics before fine-tuning begins. Cold-start
viability follows: there is nothing project-specific to calibrate, since no
target-specific ranking weights exist. The small losses observed (median
$-0.017$ MRR, worst $-0.045$ across all 63 pairs) are consistent with
similarity-based ranking degrading smoothly under a suboptimal source, though
TRANP-CNN's losses are comparably bounded (worst $-0.040$), so bounded loss is
not by itself a signature of similarity-based ranking. A non-exclusive
alternative for BLAZE's cold-start behaviour is that it reranks with the same
BGE signal that selected the candidate pool (Section~\ref{sec:threats}); the
two accounts are not separable in this study.

Two competing hypotheses warrant consideration: \emph{(i) representation
quality} (BLAZE reranks with modern frozen BGE embeddings while TRANP-CNN
and COOBA use their original designs', so its profile might reflect
representation quality rather than mechanism), and \emph{(ii) model age}
(BLAZE, 2025, is four to five years newer, so the gap might reflect general
progress), not mutually exclusive. Two observations bear on them, though
neither rules them out: the two non-BGE rerankers differ from each other in
profile ($d = 0.95$ vs.\ $0.27$), so representation vintage alone does not
determine transfer behavior; and BLAZE's CP-cold-start (0.210) trails its
CP-transfer (0.298), so the pre-trained representations alone are not
sufficient without target-side training signal. Representation quality may
still contribute to BLAZE's profile specifically, and
Section~\ref{sec:threats} records the residual representation asymmetry as a
construct threat.

\paragraph{COOBA (GNN with adversarial domain adaptation).}
COOBA trains a Gradient Reversal Layer discriminator to erase
project-specific signals from the shared GNN feature space. This
adversarial objective is the CPL mechanism, forcing the encoder toward
domain-invariant features by preventing the discriminator from
distinguishing source from target representations. But the mechanism needs
target-side representations \emph{during training} for the discriminator
to contrast against the source distribution; with zero target data it gets
no target signal, so the resulting features stay calibrated to the source
AST vocabulary (identifier namespaces, call-graph structures, type
hierarchies, all project-specific), which misalign on a new target,
explaining the near-zero cold-start MRR ($0.027$ vs.\ $0.138$ for
WP-small). The counter-intuitive finding that higher source within-project
performance predicts \emph{worse} absolute transfer is consistent with the
same logic: a model that has learned source AST patterns very precisely has
encoded more source-specific features that generalize less to a new
project's vocabulary. TRANP-CNN, which has no AST input, shows the same
negative association, however, so source-AST overfitting cannot be the whole
explanation.

\paragraph{TRANP-CNN (CNN reranker).}
TRANP-CNN produces a scalar relevance score per (bug report, candidate
file) pair via parallel convolutional feature extractors and margin
ranking loss, calibrated during training to the source project's ranking
difficulty. Without target examples this score stays anchored to the
source distribution, its scale and decision boundary not transferring,
explaining cold-start failure ($0.038$ vs.\ WP-small $0.165$). With 20\%
target labels in the same joint training run (CP-transfer,
Section~\ref{sec:scenarios}), the scalar score calibrates to the target's
ranking difficulty while still drawing on the full source signal, yielding
the largest effect of the three models ($d = 0.95$, $p < 0.001$ across all
metrics; BLAZE $0.49$, COOBA $0.27$); this jointly-trained
result is statistically equivalent to WP-large (TOST, $\pm 0.05$ MRR), the
joint signal closes the gap to the ample-data ceiling once target
calibration is provided.

A useful cross-check is model sophistication: COOBA employs a more
sophisticated adversarial mechanism than TRANP-CNN, and the two were published
a year apart, yet COOBA achieves worse outcomes across nearly all metrics. If
sophistication or recency were the explanatory variable, COOBA should at least
match TRANP-CNN; it does not, inconsistent with
publication year and complexity as the primary explanation, and more
consistent with evaluation mechanism as an explanatory dimension (an
interpretation, not a controlled result, for the reasons given in
Section~\ref{sec:models}). Specifically, whether ranking at inference time requires
target-side calibration of learned parameters appears to govern CPL
viability: models ranking via pre-trained representation similarity
(BLAZE) need no calibration and can operate in cold-start, while models
ranking via a learned scalar (TRANP-CNN, COOBA) require it and fail in
cold-start, an inference-time mechanism independent of training-time
sophistication.

\subsection{Implications for CPL Research}

\looseness=-1 The low cross-model agreement (44--54\% across the three model pairs) is a
methodological contribution for the CPL research community. It suggests that
CPL benefit may not generalize across model architectures: a finding
from one CPL model did not, in this corpus, reliably predict outcomes from a different architecture
on the same project pairs. This echoes, in a different task, the central conclusion of
the cross-project defect prediction literature introduced in Section~\ref{sec:related}:
\citet{hosseini2017systematic}'s meta-analysis found transfer success in defect
prediction to be context-dependent with no single reliably predictive factor.
Our RQ2 results partly echo this (Section~\ref{sec:results}): similarity-based
project features such as domain gap do not reliably predict pair-level gain,
and no measured feature predicts COOBA's outcomes, although target label
volume is a strong predictor for BLAZE and TRANP-CNN.

The counter-intuitive negative predictor for source quality, where better
within-project performance on the source predicts worse absolute CP-transfer
MRR, appears for both COOBA and TRANP-CNN (Section~\ref{sec:results}), so it
is not specific to COOBA's AST-based adversarial design; for both models it
does not extend to gain over WP-small. It suggests that within-project
performance and cross-project generalizability may be competing objectives for
models that learn source-specific scoring functions, a mechanism we did not
test. This motivates CPL-specific model design that explicitly prioritizes
cross-project robustness over within-project peak performance, a direction we
are not aware of existing proposals addressing. The pattern does not appear
for BLAZE ($\rho = -0.02$), whose frozen backbone limits how far training can
specialize to the source.

Beyond architectural choices, this study yields several observations about how
future CPL evaluations should be designed, summarized in Section~\ref{sec:conclusion}:
reporting against a limited (not ample-data) within-project baseline, matching
the target label budget between conditions, and reporting negative-transfer rates and
zero-label cold-start viability alongside win rates and mean gains. The cold-start
result for BLAZE is among the most practically useful findings in this study, yet it
is invisible under a standard fine-tuning-only evaluation protocol.

\subsection{Threats to Validity}
\label{sec:threats}

\paragraph{Internal validity.}
All three models depend on FAISS retrieval at the first stage. Recall at
$K = 300$ varies across the 13 projects from 31.3\% (ansible) to 100\%
(xarray), median $\approx$70\%; four projects fall below 50\% (ansible
31.3\%, wagtail 34.6\%, meson 45.5\%, jupyterlab 47.0\%), where the reranker never sees the
ground-truth file in roughly half to two-thirds of queries. This imposes a
shared performance ceiling that suppresses absolute metrics and conflates,
in Table~\ref{tab:stratified}'s large-codebase degradation, model limitation
with retrieval ceiling. Disentangling the two would require
re-ranking-only evaluation over gold-standard candidate sets, left to future
work. The within-model CPL comparison (CPT vs.\ WPS) is unaffected, since
both conditions face the same constraint.

\textbf{Training stochasticity.}
The fixed seed (42) controls the data splits, not the training process:
weight initialization, data-loader ordering, and non-deterministic GPU
kernels are unseeded, so identically configured runs differ. Every scenario is
reported from a single run because repeating the design is computationally
prohibitive: one pass over all 606 training runs took five to six months of
GPU time (Section~\ref{sec:dataset}), so three repetitions per scenario would
take well over a year. \emph{Individual}-pair gains therefore carry run-to-run
variation that this study does not quantify, and pair-level results, including
the illustrative cases (Section~\ref{sec:cases}) and the classification of
individual pairs as negative transfer, should be read with that in mind; the
paired Wilcoxon analysis over 63 pairs is less exposed to it than any single
pair. For the same reason, the $\pm 0.01$ MRR band used to classify negative
transfer is a conventional ``any meaningful loss'' threshold rather than one
calibrated to run-to-run variance.

\textbf{Random rather than chronological splits.}
Splits are drawn at random (seed 42) rather than chronologically, which can
leak future information: a test bug's training split may include later-filed
bugs touching the same file, exposing a fix pattern that would not have
existed at deployment time. We did not re-run the pipeline chronologically
for this study; we flag it as an unaddressed threat and recommend a
chronological sensitivity check before treating the random-split results as
an upper bound on realistic deployment performance.

\paragraph{Construct validity.}
MRR and Top-$K$ measure file-level localization; finer-grained (function- or
line-level) transfer patterns are untested here, though the evaluation is
aligned with prior CPL benchmarks.

\textbf{Input-representation asymmetry.}
BLAZE reranks with modern frozen BGE embeddings (adapter-only fine-tuning);
TRANP-CNN trains token embeddings from scratch and COOBA uses GloVe-300,
both faithful to their original designs. Representation quality differs
across rerankers and is bundled into each architecture rather than
controlled out. This is bounded by three observations: the candidate pool
is identical for all models; the two non-BGE rerankers differ from each
other ($d = 0.95$ vs.\ $0.27$), so vintage alone does not determine CPL
profile; and the cold-start dichotomy is consistent with the mechanism
account in Section~\ref{sec:architecture}. A more specific asymmetry: the shared
FAISS stage pools candidates via mean-pooled BGE embeddings, and BLAZE
reranks that same pool with MaxSim over the same signal that selected it,
while TRANP-CNN and COOBA apply an independently-learned score to a pool
chosen by a representation they never see. BLAZE thus re-scores its own
retrieval signal with better aggregation rather than an independent
signal, a non-exclusive alternative to the evaluation-mechanism account of
its cold-start advantage (Section~\ref{sec:architecture}) that this study
cannot separate from it; an untrained BGE+MaxSim baseline would isolate this
aggregation effect from adapter training and is the most valuable follow-up
addition.

\textbf{Re-implementation fidelity.}
TRANP-CNN and COOBA were re-implemented for this study
(Section~\ref{sec:models}); loss functions, encoders, and adaptation
mechanisms follow the originals, but the shared FAISS stage and, for
TRANP-CNN, the BGE tokenizer are departures, so absolute scores are not
comparable to the original papers (within-study comparisons are
unaffected). We cannot fully exclude COOBA's weak profile reflecting our
re-implementation rather than the adversarial paradigm itself. We verified
it against the original design equation-by-equation (bug encoder, feature
extractors, gradient-reversal discriminator, margin-ranking loss) and found
it faithful on every core mechanism, with two deliberate departures:
source and target task losses are summed in the
same step rather than alternating per the original's Eq.~12, and evaluation
uses one held-out split rather than 10-fold cross-validation. Re-running
this re-implementation on the original paper's own benchmark data under a
closer protocol is a scoped follow-up we did not carry out; until then the
weak COOBA result should be attributed to the adversarial paradigm only
provisionally. No hyperparameter search was run for any model; each uses
the values reported in its original paper, including COOBA's
$\lambda_\mathrm{adv} = 0.1$ (Table~\ref{tab:hyperparams}). COOBA's profile
may therefore partly reflect an untuned adversarial weight rather than the
paradigm itself, a sensitivity we did not test.

\textbf{Pre-training data leakage.}
\texttt{BAAI/bge-code-v1} was pre-trained on a corpus that very likely
includes these popular repositories~\citep{aleithan2024swebenchplus}, which
may inflate BLAZE's cold-start result via memorized project-specific
representations. A proxy check using GitHub star counts shows no significant
Spearman correlation with cold-start MRR ($|\rho| < 0.20$), suggesting
familiarity is not the primary driver, though a definitive test needs
repositories absent from pre-training corpora. Separately, tangled commits
(alongside the $>$10-file patch filter applied during curation) introduce
residual label noise as a construct threat.

\textbf{Domain classification.}
As noted in Section~\ref{sec:dataset}, the four domain labels used throughout
RQ2 (Table~\ref{tab:crossdomain}) were independently assigned by two raters
who agreed on 11 of 13 projects (84.6\% raw agreement), disagreements resolved by the second author.
This is raw percent agreement, not a chance-corrected statistic such as
Cohen's $\kappa$; with four unevenly sized categories, raw agreement
can overstate reliability relative to $\kappa$, so this should be read as a
partial, not complete, mitigation of the labeling threat.

\textbf{Empirical holdout validation is itself a data requirement.}
Recommending a small validation holdout (Section~\ref{sec:discussion}) is in
tension with the zero-label framing motivating this study: a project able to
construct one is not fully in that regime. We lack a principled minimum
size; as a rough guide the smallest WP-small set here (24 bugs,
\texttt{pydata/xarray}) still yielded a usable model, suggesting holdout
validation becomes practical around a similar order of magnitude, roughly two
dozen bugs, rather than at zero labels. Training and holdout sets
serve different purposes, though, so this is only a rough analogy.

\paragraph{External validity.}
The study covers 13 Python projects; generalizability to other languages,
proprietary codebases, or different bug-report styles is unvalidated. The
Python scope is by design. The mechanism-based explanations
(Section~\ref{sec:architecture}) are grounded in inference-time properties
that are language-agnostic in principle, but COOBA's AST-based encoder is
sensitive to identifier namespaces, call-graph depth, and type hierarchies
that differ between Python and Java, so whether its failure mode replicates
there is open. Only open-source projects with public commit histories are
included, and the embedding backbone is specific to its evaluated version; a
different backbone may shift absolute numbers while leaving relative
conclusions intact. These constraints compound: the corpus favors popular,
well-instrumented repositories that are also likely in the backbone's
pre-training data, a combined selection effect to weigh when generalizing
absolute figures (the within-model CPT-vs-WPS comparisons are less exposed).

\textbf{No LLM-agent comparison.} The models studied here are not compared
with LLM-agent fault localization (Section~\ref{sec:related}): agents are
evaluated on a different task and metric (patch generation over a shortlist),
and no agent baseline is available at the Top-$K$/MRR/MAP metrics reported.

\paragraph{Conclusion validity.}
The cross-model consistency finding rests on pairs shared across all three
models. Wilcoxon tests are conducted independently per metric with
Bonferroni controlling the familywise error rate per five-metric family. The
63-pair set is structured rather than random, over-representing
DS$\times$DS pairs (Group A, 30 pairs); the RQ2 stratified analysis
(Tables~\ref{tab:stratified}, \ref{tab:crossdomain}) isolates within- from
cross-domain effects. The 13 target projects recur unevenly (3--12 times
each, median 3, mean 4.8; jupyterlab most often at 12, its Group~B pivot
role), so pairs are not independent observations. The primary RQ1 Wilcoxon
tests (Section~\ref{sec:results}) treat all 63 pairs as independent; we
additionally fit linear mixed-effects models of CPL gain with target
project as a random intercept, following recent guidance
for clustered observations~\citep{oliveira2019evolution}
(\texttt{Scripts/analysis/mixed\_effects\_check.py}). Clustering matters:
between-project variance ($0.0013$) is about two-thirds of the residual
($0.0020$; intraclass correlation $\approx 0.39$). In a pooled model with
BLAZE as reference, TRANP-CNN's gain is significantly higher ($+0.020$,
$p = 0.012$) while COOBA's does not differ significantly ($-0.011$,
$p = 0.17$): the clustered analysis separates TRANP-CNN from the other two.
Fitted per model, the mean CPL gain is significantly positive only for
TRANP-CNN ($+0.041$, $p = 0.001$) and not for BLAZE ($+0.017$, $p = 0.18$) or
COOBA ($+0.009$, $p = 0.47$); a Wilcoxon test on the 13 per-target mean gains
gives the same picture (TRANP-CNN $p = 0.007$, 10/13 targets positive; BLAZE
$p = 0.27$, 7/13; COOBA $p = 0.32$, 6/13). BLAZE's and COOBA's pair-level
significance in Table~\ref{tab:wilcoxon} therefore depends on treating pairs
sharing a target as independent and should be read as suggestive, partly a
power limitation at 13 clusters; TRANP-CNN's benefit is robust to this
clustering, though with only 13 clusters all of these estimates should be read
cautiously.

\section{Conclusion}
\label{sec:conclusion}

Bug localization models require labeled training data that newly started or
data-scarce projects often lack, and CPL has been proposed to close that gap,
but to our knowledge no prior work has compared it against the practically
relevant within-project-limited baseline or tested whether its findings
generalize across model implementations. Every published CPL paper we are
aware of evaluates one model against an ample-data comparator its intended
users cannot reach. We addressed both gaps with a controlled study across 63
source$\to$target pairs, 13 Python projects, and three architecturally
distinct CPL models (embedding-based, convolutional, adversarial-GNN) under
four scenarios isolating the contribution of cross-project knowledge.

Cross-project transfer outperforms within-project limited training for all
three models at the pair level, but by very different amounts: a large effect
for TRANP-CNN ($d = 0.95$), robust to clustering pairs by target project; a
smaller one for BLAZE ($d = 0.49$), not significant once pairs are clustered;
and a small one for COOBA ($d = 0.27$), our only adversarial GNN instance and
one we re-implemented, which does not survive multiple-comparison correction
and collapses to near-zero in cold-start (Section~\ref{sec:results}). BLAZE's
zero-label deployment retains about 78\% of WP-small's MRR and is
viable in nearly half of pairs;
TRANP-CNN reaches full within-project performance with 20\% target labels, the
highest CPL ceiling relative to its own baseline, though BLAZE still leads on
absolute MRR and Top-K throughout. CPL gain correlates strongly with how few
labels the target already has: the least-labeled targets, the ones this
study is written for, benefit the most. This benefit has limits, however:
negative transfer affects 30.2\% of BLAZE's and COOBA's deployments and 9.5\%
of TRANP-CNN's, common for two of the three architectures and present for
all three; domain sensitivity is pronounced for BLAZE and modest for the
other two; similarity-based project features do not reliably predict pair-level
success, though target label count does; and cross-model direction agreement
ranges from only 44\% to 54\%, so a result from one architecture does not
reliably predict another's even on the same pair, though the models' gains are
positively correlated (Section~\ref{sec:discussion}).

For practitioners, BLAZE's zero-label deployment reaches the viability
threshold in nearly half of pairs, and once a modest label set is added its
CP-transfer beats WP-small in 63.3\% of within-domain pairs; TRANP-CNN needs 20\% of target labels to match full
within-project training, and gains the most from cross-project data when it
has them. Both carry a cost WP-small does not: CP-transfer
trains on the full source corpus in addition to the target-side labels, and neither guarantees a win. Source and domain alignment should be checked empirically, not
assumed, particularly for BLAZE.

Future work should pursue: an untrained BGE-with-MaxSim
baseline to separate the aggregation effect from adapter training, which
requires no model training; extending to multi-language CPL (Java,
cross-language pairs, Kotlin/Go) to
test whether these findings generalize beyond Python~\citep{yang2024multi};
model-specific source-selection criteria beyond project metadata; and
function- or line-level localization.

The most durable contribution of this study, beyond any single number, is
methodological: low cross-model agreement suggests CPL benefit may not
generalize across architectures, so future CPL evaluations should report
results per-architecture, against a limited (not ample-data) baseline, with the
target label budget \emph{verified as} matched between conditions, and with negative-transfer rates reported
alongside win rates and mean gains.

\section*{CRediT authorship contribution statement}
\textbf{A Eashaan Rao:} Conceptualization, Methodology, Software, Data curation,
Investigation, Formal analysis, Visualization, Writing -- original draft.
\textbf{Sridhar Chimalakonda:} Conceptualization, Methodology, Supervision,
Writing -- review \& editing.

\section*{Declaration of competing interest}
The authors declare that they have no known competing financial interests or
personal relationships that could have appeared to influence the work reported in
this paper.

\section*{Funding}
This work was supported by the Indian Institute of Technology Tirupati. The
funder had no role in study design, data collection and analysis, decision to
publish, or preparation of the manuscript.

\section*{Declaration of generative AI and AI-assisted technologies in the writing process}
During the preparation of this work, the authors used AI-assisted tools (RISHA
Lab, IIT Tirupati) to improve the clarity, presentation, and language of the
writing. After using this tool, the authors reviewed and edited the content as
needed and take full responsibility for the content of the publication. AI
tools were not used to generate the study's research ideas, experimental
design, analysis, or conclusions.

\section*{Acknowledgements}
We thank Ajinkya Savarkar for independently classifying the project domains
reported in Section~\ref{sec:dataset}.

\section*{Data Availability}

The replication package for this study is publicly archived on Zenodo at
\url{https://doi.org/10.5281/zenodo.21560518}.
It includes the curated bug metadata databases for all projects, experiment result CSVs, the four-scenario evaluation pipeline, and all analysis scripts. The package supports full reproduction of Tables~\ref{tab:scenario_means}--\ref{tab:source_selection}
and all figures in Section~\ref{sec:results}. Given the re-implementation
fidelity threat discussed in Section~\ref{sec:threats}, the package explicitly
includes our COOBA and TRANP-CNN re-implementations (source, configuration,
and the equation-by-equation verification notes against the original
publications), alongside our adaptation of BLAZE, so that the
re-implemented components can be independently audited. The ACM SIGSOFT
empirical standards checklist accompanying this study (Section~\ref{sec:design})
is also included in the package.

\bibliographystyle{cas-model2-names}
\bibliography{cas-refs}

\end{document}